%% file: main.tex
\documentclass[12pt]{article}
\usepackage[utf8]{inputenc}

\usepackage{amsmath,amsthm,color,amssymb,fullpage,setspace,nicefrac,natbib,comment,mathtools,subfig}
\usepackage{dsfont}
\usepackage{indentfirst}
\usepackage{soul}
\usepackage{url} 
\usepackage[colorlinks=true,linkcolor=blue, citecolor=blue]{hyperref}

\usepackage[capitalize,noabbrev]{cleveref}

\setcitestyle{citesep={,}}

\usepackage[textsize=footnotesize]{todonotes}

\newcommand{\elaine}[1]{{\color{magenta} [{elaine:} #1]}}
\newcommand{\jacob}[1]{{\color{blue} [{jacob:} #1]}}
\newcommand{\ignore}[1]{{\color{lightgray}[#1]}}
\renewcommand{\elaine}[1]{}
\renewcommand{\jacob}[1]{}
\renewcommand{\ignore}[1]{}
\setuptodonotes{disable}

\newtheorem{prop}{Proposition}
\newtheorem{theorem}{Theorem}
\newtheorem{lem}{Lemma}
\newtheorem{remark}{Remark}
\newtheorem{definition}{Definition}
\newtheorem{corollary}{Corollary}
\newtheorem{thm}{Theorem}[section]      

\newtheorem{lemma}[thm]{Lemma}

\newcommand{\ledger}{{\sf Log}}
\newcommand{\nodes}{N}
\newcommand{\tx}{\sf tx}
\newcommand{\chain}{\sf Chain}
\newcommand{\B}{{\sf B}} 

\usepackage{enumitem}

\title{On the Viability of Open-Source Financial Rails: \\Economic Security of Permissionless Consensus\\
{\large see \href{https://arxiv.org/abs/2409.08951}{here} for the latest version }
}

\author{Jacob D. Leshno\footnote{University of Chicago Booth School of Business, Jacob.Leshno@ChicagoBooth.edu.  }
\and Rafael Pass\footnote{Tel Aviv University and Cornell Tech, Computer Science, rafaelp@tau.ac.il.}
\and Elaine Shi\footnote{Carnegie Mellon University, Computer Science and Electrical and Computer Engineering, runting@gmail.com.}
}

\date{\today\footnote{
This work is supported by NSF grant number 2212747 and by the Robert H. Topel Faculty Research Fund at the University of Chicago Booth School of Business.
We thank Srivatsan Sridhar and Dionyziz Zindros for their insightful and detailed comments. 
We thank Ittai Abraham,
Garth Baughman,
Dirk Bergemann,
Eric Budish, 
Francesca Carapella,
Alex Frankel,
Rod Garratt,
Niels Gormsen,
Emir Kamenica,
Anil Kashyap,
Andrew Lewis-Pye,
Yueran Ma,
Dahlia Malkhi,
Harry Pei,
Tim Roughgarden, 
Marciano Siniscalchi,
Ertem Nusret Tas,
David Tse,
Maarten van Oordt,
Luigi Zingales,
and seminar participants at 
the ACM SIGecom Winter Meeting,
the Federal Reserve Board of Governors, Market Design NBER, a16z Research Lab, Stanford, Princeton, 
the Northwestern Conference on the Interface between Computer Science and Economics, 
the London Symposium on Information Theory, and the Simons Institute for the Theory of Computing. All errors are our own. 
}
}

\begin{document}
\maketitle

\begin{abstract}
Bitcoin demonstrated the possibility of a financial ledger that operates without the need for a trusted central authority. 
However, concerns persist regarding its security and considerable energy consumption.
We assess the consensus protocols that underpin Bitcoin’s functionality, questioning whether they can ensure economically meaningful security while maintaining a permissionless design that allows free entry of operators.
We answer this affirmatively by constructing a protocol that guarantees economic security and preserves Bitcoin's permissionless design. 
This protocol's security does not depend on monetary payments to miners or immense electricity consumption, which our analysis suggests are ineffective. 
Our framework integrates economic theory with distributed systems theory, and formalizes the role of the protocol's user community. 

\end{abstract}

\input{1_Intro}

\input{2_Model_and_Def}

\input{3_Bitcoin_Analysis}

\input{4_StubNaka_3}

\input{5_PoS}

\input{6_Discussion}

\bibliographystyle{apalike}

\bibliography{_DistEconSecurity,_FinanceTrust}

\appendix


\input{A1_Appendix}

\input{A2_what_deters_attacks}

\end{document}

%% file: 1_Intro.tex
\newpage
\section{Introduction}

\jacob{\\
This paper makes the following contributions:

-- A formal framework for what the protocol should do (record keeping), and what does it mean for it to be secure (economic security)

-- Argue that the problems with Bitcoin's incentive are inherent to its reliance on a large pool of anonymous nodes

-- Show that despite these challenges, it is possible to create a secure permissionless protocol

-- Provide a distributed systems model that formally incorporates the reliance on the "community" or social factors. \\
\hspace*{2cm}    ---- allowing discussion of the "social consensus" layer\\
\hspace*{2cm}    ---- avoiding impossibility results (that are unrealistic)

}

\jacob{
SN designed to be safe, resilient to attacks, and minimize dependence on any centralized or external entities.
}

\jacob{
=====================
Why did I do this paper? 

We need to understand what these systems do -- record keeping -- and what properties should we set for them.

It's important to think about "common knowledge", what can the protocol/communication give, what is needed from the "community", and what market structure or controlling power can we have. 

The results are important because they show the possibility of an open and secure system. 

The results also explain the current system: maybe we need a wider definition of the current Bitcoin protocol. 
=====================
}

\jacob{An important distinction from common knowledge (see Yannai's paper). }

\jacob{
Blockchain = record keeping. 
That's a really difficult thing, and sufficient for payment systems
(other things can be built on top of it)
}

\jacob{
Bitcoin security does not come from the mining expenditure, "trust" is not related to the power to produce a block. 
}

\jacob{Add:

\url{https://www.nytimes.com/2024/09/24/business/dealbook/visa-antitrust-lawsuit-debit-cards.html}

\url{https://www.axios.com/2024/09/24/visa-antitrust-debit-card-monopoly-lawsuit}
}

\jacob{explain: "honest" == agreement between parties who want to agree. Can't do more. }

\jacob{explain: We don't do eq because we don't specify payoffs. We give the game form, i.e., what players can do by deviating. Thus, we're more general. }

A long-standing literature has demonstrated the critical role of the financial system in fostering economic growth \citep{schumpeter1911theory, rajan1998financial}, and the importance of a trusted financial system that protects against fraud and deception \citep{la2006works, guiso2004role, guiso2008trusting}. 
While acknowledging its benefits, economists have raised concerns regarding the current state of the financial system, particularly highlighting inefficiencies and rent-seeking behavior \citep{greenwood2013growth, admati2014bankers, philippon2015has}. 
Payment-processing firms, the operators of the rails of the financial system, collected global revenues of $2.2$ trillion in 2022 \citep{McKinseyPayments2023}, and are frequently subject to regulatory scrutiny and antitrust litigation aimed at curbing their market power.\footnote{
Recent high-profile cases include the U.S. vs. American Express \citep{us_v_american_express_2010}, the European Commission imposing interchange fee regulation \citep{EUInterchangeFee2015}, and the DOJ blocking Visa's acquisition of Plaid \citep{DOJVisaPlaid2020}. 
For lists of antitrust lawsuits and regulatory actions, see Table 5 in \citet{herkenhoff2020bears} and \citet{hayashi2019public}, respectively. 
}

\cite{satoshi-bitcoin} showed the possibility of an open-source alternative.
It introduced Bitcoin, the first functioning payment system\footnote{
Bitcoin offers the basic functionality of a payment system. Ownership of funds is verified through cryptography. Standard accounting principles, such as ensuring that each transaction's credits equal its debits, are enforced through an open and auditable ledger.
}
operated by anonymous interchangeable computer servers (known as nodes or miners) that can freely enter and exit the network. 
Bitcoin's economic design, particularly the free entry of its miners, shields users from monopoly pricing and eliminates the pricing power of the system's operators.\footnote{\cite{huberman2021monopoly} show that the free entry of miners in Bitcoin's Nakamoto protocol prevents any miner (even large ones) from profitably affecting transaction fees. Empirical work by \cite{easley2019mining} and \cite{donmez2022transaction} supports these findings.}
This economic design was made possible by a technological breakthrough: the Nakamoto permissionless consensus protocol, which coordinates a network of untrusted participants who collectively maintain a distributed ledger. In other words, the Nakamoto consensus protocol creates a record-keeping service without a trusted recordkeeper.
The potential of such ``decentralized'' consensus protocols sparked significant public interest and extensive research. 
Recently, the Web3 movement aspired to leverage this technology to curtail the market power of Big Tech platforms and develop open-source alternatives.\footnote{See 
\cite{dixon2024read} for an exposition of these ideas and historical precedents, including the success of the open-source operating system Linux and the community-controlled Domain Name System (the internet’s ``phone book'').
For media coverage, see \url{https://www.wired.com/story/web3-paradise-crypto-arcade/}
and \url{https://www.wired.com/story/how-blockchain-can-wrest-the-internet-from-corporations/}
(accessed August 2024).
}

\jacob{
Free are feeds and web3 efforts are certainly good and have an admirable purpose, but their roadmap is unclear. 
What do people mean by "decentralization"? Which properties should these systems hold? What are the services such systems can and cannot provide? (record keeping is easy, enforcing property rights is hard)
}

Yet, despite the significant attention cryptocurrencies have garnered, mainstream adoption remains limited \citep{bis2023crypto}.\footnote{\cite{makarov2021blockchain} find that much of cryptocurrency user activity has been linked to speculative behavior or black market transactions. \cite{alvarez2023cryptocurrencies} find limited adoption for payments. \cite{auer2023crypto} analyze Bitcoin adoption and find that financial speculation drives entry of new users. \cite{yermack2019blockchain} evaluates the technological promise but concludes that ``significant disruptions to the legacy financial system still appear to be years away.''
Critiques in the press are common, see, for example, \url{https://www.forbes.com/sites/yayafanusie/2021/01/01/stop-saying-you-want-to-bank-the-unbanked/}, and \url{https://www.wired.com/story/whats-blockchain-good-for-not-much/}. }
The issue extends beyond the adoption challenges faced by any new technology. There are fundamental questions about Bitcoin's design and the viability of any permissionless consensus protocols for mainstream financial applications. In particular, 
Bitcoin's Nakamoto protocol is vulnerable to corrupt majority attacks (also called $51\%$ attacks or double-spending).\footnote{In a corrupt majority attack, a malicious actor can create a manipulated ledger (which reverts transactions) and exploit the consensus protocol to get all others to adopt the manipulated ledger (see \Cref{sec:Bitcoin_incentives}).}
This vulnerability has attracted much attention from practitioners, economists, and computer scientists, questioning whether permissionless consensus protocols can be made secure at a reasonable cost.\footnote{\cite{budish2022economic} critiques Bitcoin's Nakamoto protocol for its vulnerability to double-spend attacks. \cite{shanaev2019cryptocurrency} provides a list of successful double-spend attacks of blockchains that utilize the Nakamoto protocol. See the literature review below for further details.}



We approach this problem as an economic design problem, asking whether it is possible to design a protocol that has the free entry properties that are provided by a permissionless design while also providing a security guarantee that is economically meaningful.

In this paper, we theoretically evaluate whether consensus protocols can provide economically meaningful security while preserving a permissionless design that allows for the free entry of operators.
Our primary finding is that a permissionless and economically secure protocol is indeed feasible. We demonstrate this feasibility by constructing a secure modification of Bitcoin's Nakamoto protocol that maintains its permissionless design. Notably, our protocol adopts a different approach to security; its security does not depend on monetary incentives for miners or their excessive energy consumption, which we identify as ineffective.




 


To perform our analysis, we use a framework that integrates economic theory with distributed systems theory, both components being necessary and complementary. The distributed systems model is essential for analyzing general protocols, user behaviors, and attacks. Additionally, economic incentives depend on the system's ability to detect and punish deviations from prescribed behaviors. The distributed systems model enables us to assess the knowledge of each participant and determine whether an attack is detectable. 
The economic model allows us to account for attacks that operate or incentivize nodes within the system, and it provides a meaningful quantification of such security vulnerabilities in terms of the cost of the attack.


We present a definition of \emph{economic security} for general consensus protocols that captures whether users of the system are protected against deception or fraud in the face of general attacks. 
Consider a merchant who wishes to receive payment. The merchant wants to ensure that the payment is valid, i.e., the credit is recognized by others and can be spent.
In a digital payment system, the merchant relies on a payment terminal to determine whether the payment is valid. 
In distributed systems terminology, the merchant problem is achieving consensus on the validity of the payment. 
If the distributed systems property \emph{consistency} is satisfied, the merchant's terminal is guaranteed to be in agreement with the rest of the network, and the merchant cannot be deceived. Therefore, we define the economic security of a distributed ledger as the economic cost required to violate consistency (Section \ref{sec:definition} provides background and motivation for the definition). 
This definition captures the corrupt majority attack mentioned above,\footnote{A double-spend attack that deceives the merchant constitutes a consistency violation.} as well as other attacks.\footnote{For example, an attacker can potentially profit from rewriting old parts of the ledger to ``retroactively'' buy an option after its expiration. See \cref{sec:ex_profitable_attacks} for further detail.}



We analyze Bitcoin's Nakamoto protocol to highlight challenges in securing permissionless protocols.
We evaluate the economic security of Bitcoin's Nakamoto protocol in two economic environments, selected to underscore the deficiencies in incentives for securing permissionless protocols.
Two shortcomings compromise the system's security, regardless of the level of mining rewards. 
First, the protocol fails to detect attacks. Malicious miners can generate an alternative ledger and cause the system to adopt it, deleting transactions and violating consistency. 
The protocol does not recognize that this alternative ledger is a product of malicious behavior, treating these miners the same as honest ones (we explain the protocol's rationale in \cref{sec:explaining_bitcoin_faults}). 
In particular, the protocol fails to punish such attackers and does not provide incentives to disincentivize attacks.
Second, even if attacks prompt an external response, such as a collapse in the exchange rate, the resulting harm affects all miners, irrespective of their involvement in the attack. 
If miners are small, the tragedy of the commons can arise.
Together, these deficiencies underscore a broader challenge in relying on incentives to secure permissionless protocols.


The absence of attacks on Bitcoin and Ethereum so far suggests that other factors deter attacks. 
We highlight one in particular: the community response. 
Recall that the protocol is a record keeper that maintains a ledger. An attack can cause the nodes to adopt a corrupted ledger.
However, if the community detects the ledger is corrupted, they may manually override the nodes to restore the correct ledger.  
This community response is powerful, as it effectively nullifies the attack and serves as a deterrent against future attacks. 
However, this response requires that the community be able to detect attacks and maintain agreement on the correct ledger. 

The community's response suggests an alternative approach to protocol security. Rather than relying solely on incentives and monetary rewards, a protocol could achieve security by automating and integrating a community-like response into its design.
That is, the protocol should maintain agreement on the correct ledger. 
Instead of allowing attacks to corrupt the ledger and subsequently restoring it, the protocol should enable all nodes to preserve the correct state and ignore attacks that attempt to corrupt it. 
By considering a version of the community response within a formal protocol, we can develop alternative protocols and rigorously evaluate the capabilities of the community. 

We demonstrate the feasibility of a permissionless consensus protocol that is economically secure by constructing such a protocol.  
Our protocol, which we name Stubborn Nakamoto, is a modification of Bitcoin's Nakamoto protocol. 
It is designed for the same environment as the Nakamoto protocol (namely, synchrony with late joiners). 
In the absence of an attack, the Stubborn Nakamoto protocol and the Nakamoto protocol operate essentially identically.
Stubborn Nakamoto preserves the free entry of nodes. All nodes serve equivalent roles, remain anonymous, and can join or leave at any time. 
We prove that the Stubborn Nakamoto protocol is economically secure, showing that it guarantees consistency even against an attacker with unlimited computational power and economic resources. In particular, corrupt majority attacks (such as double-spending) are rendered impossible under the Stubborn Nakamoto protocol. 

The Stubborn Nakamoto protocol differs from the standard Nakamoto protocol in two key aspects. First, it specifies an explicit condition under which a node finalizes a record in its ledger. Once a record is finalized, the node never reverts it. Second, the Stubborn Nakamoto protocol instructs nodes to wait before finalizing a record. This delay allows nodes to determine whether any other node observes an inconsistent ledger and ensures that the ledger is updated only when the node is certain that no conflict is possible. Together, these measures ensure that if a node finalizes a record, all other nodes are in agreement. 

Some trade-offs in protocol design are inevitable. Impossibility results show that no protocol can make progress  (i.e., add records to the ledger) and guarantee security when attacked by a sufficiently powerful adversary.\footnote{See \cite{pass2017rethinking,lewis2023permissionless}, and \cref{sec:SN_recovery}.} Our Stubborn Nakamoto protocol prioritizes security. The protocol is able to detect whether it is impossible to securely make progress, and halts progress to guarantee security. We show that the Stubborn Nakamoto protocol can recover from attacks with external help from the community, which we model as a recovery oracle. We prove that such external help is necessary and show that the required recovery oracle can be plausibly implemented. 

Our results show the feasibility of an economically secure ledger that operates without the need for a trusted record keeper.\footnote{
This paper focuses on the technology's ability to offer a record-keeping service, that is, to maintain a ledger, as record-keeping is a core service of any financial system, and payment systems in particular (famously,  \cite{kocherlakota1998money} reasons that ``Money is Memory''). See \cref{sec:Discussion} for a discussion of additional challenges. 
} It protects its users from monopoly's harm, allows free entry and exit of its operators, and renders its operators exchangeable. \jacob{Users don't need to be anonymous}
The analysis offers an alternative explanation for Bitcoin's security, suggesting a broader perspective that includes the role of the community. Attacks that deceive the protocol may be rendered ineffective if they can be detected and reversed by the community. The security of a ledger, therefore, may have little to do with the amount of resources spent on it. Instead, security is ensured by establishing agreement among all participants.


\jacob{Side note: it's about the knowledge, not the per se mining.
Correct misconceptions about the longest chain rule. 

For example, Bitcoin does not "erase" Bitcoin cash. 
}

\jacob{comments about PoS?}\todo{PoS?}




\subsection{Related work}
This paper builds on work in economics and in distributed systems, bringing together elements of both.  

We build on a rich literature in economics that studies double-spend attacks on Bitcoin's Nakamoto protocol, including \cite{budish2022economic}, \cite{auer2019beyond}, \cite{bonneau2016buy}, \cite{chiu2022economics}, \cite{moroz2020double}, \cite{garratt2023fixed}, \cite{pagnotta2022decentralizing}, and \cite{gans2024zero}. 
This literature focuses on the double-spend attack against Bitcoin's Nakamoto protocol. 
Our framework allows us to give a definition of economic security that considers general attacks and to analyze alternative protocol designs. 
\cite{abraham2006distributed}, \cite{abadi2022blockchain}, \cite{garay2013rational}, \cite{biais2019blockchain}, \cite{badertscher2021rational}, and \cite{halaburda2021economic} offer economic models of consensus protocols. See \cite{halaburda2022microeconomics} for a review. 

\cite{eyal2018majority} demonstrate that Bitcoin miners can earn higher rewards through strategic behavior. A significant body of literature followed exploring strategic actions in blockchain protocols, including \cite{kiayias2016blockchain, carlsten2016instability, cai2024profitable}. Such manipulations do not compromise the protocol's consistency,\footnote{Because the Nakamoto protocol guarantees consistency under honest majority, deviation by small miners cannot lead to consistency violations.} and are therefore beyond of the scope of this paper.

Our work builds on the seminal work in distributed systems. We adopt the modeling approach established in this literature in works such as \cite{pease1980reaching, lamport1982byzantine,dwork1988consensus,fischer1985impossibility}. 
Our work also builds on more recent developments in distributed systems that provide rigorous analysis of the Nakamoto protocol \citep{garay2015bitcoin, pass2017analysis,guo2022bitcoin}. 

We diverge from this literature in considering an economic notion of security and free entry of nodes. 
Classic consensus protocols, such as Paxos \citep{leslie1998part}, are designed for consensus among a fixed set of known nodes (e.g., a firm's server and its backups).
Their goal is to provide resilience to faults such as hardware or software failure and electricity outages. In such a closed deployment scenario, incentives are a non-issue.  
These protocols can lose consistency if a sufficient number of nodes are corrupted, as such events are rare in the settings they were designed for. 
By contrast, in environments with free entry, the number of corrupt nodes is endogenous; an attacker willing to bear the cost can control as many nodes as necessary to execute an attack. Therefore, our Stubborn Nakamoto protocol is designed to remain secure regardless of the number of corrupt nodes.

Free entry imposes an additional challenge: nodes may join the protocol at any point in time. There are classic consensus protocols, such as Dolev--Strong \citep{dolev1983authenticated}, that are designed for a fixed set of nodes and can operate securely (i.e., guarantee consistency and liveness) given any number of corrupt nodes. However, if nodes can join at any point and there are sufficiently many corrupt nodes, then any protocol, including Dolev--Strong, can lose consistency. Our research builds on the works of \cite{lewis2023permissionless}, \cite{budish2024economic}, and \cite{pass2017rethinking} that prove this impossibility result and show additional difficulties that arise in the permissionless setting.

The analysis of bribery attacks on Bitcoin builds upon the ideas introduced in \cite{bonneau2016buy}. Contemporaneously and independently of this work, \cite{newman2023decentralization} also argues that bribery attacks are cheap when miners are small and rational because small miners are unlikely to be pivotal to an attack's success.

Earlier works suggested using irreversible ``checkpoints'' to secure blockchain protocols \citep{snowwhitepos, Casper,checkpoint-aggelos,buterin2017casper, ebbandflow, weaksubjective, sankagiri2021blockchain}. In contrast to previous approaches that rely on adding finality gadgets or beacons that are external to the protocol, our Stubborn Nakamoto directs nodes to be ``stubborn'' given their local record, and it is provably secure without relying on any external support.  

While we focus on protocols that allow fully anonymous nodes (commonly called proof-of-work), it is worth noting that some proof-of-stake consensus protocols can provide Byzantine forensics and accountability~\citep{bftforensics,accountabilitydilemma}. In other words, if nodes misbehave, it may be possible to produce cryptographic evidence that implicates them. Such cryptographic forensics evidence can serve as a deterrent to attacks on these proof-of-stake protocols. However, \cite{budish2024economic} show the limitations of this approach. 

Finally, this work is related to \cite{sridhar2023better}, which is concurrent with an earlier draft of the present paper. \cite{sridhar2023better} demonstrate that a range of proof-of-stake consensus protocols can be modified to guarantee consistency even under a corrupt majority and suggest a recovery approach following attacks. A key difference is that our Stubborn Nakamoto protocol does not require the setup or infrastructure of proof-of-stake protocols. 

\subsection{Paper outline}
The remainder of the paper is organized as follows. 
\Cref{sec:definition} provides our model and motivates our definition of economic security.  
\Cref{sec:Bitcoin_incentives} analyzes Nakamoto's economic security and shows deficiencies in its incentives. 
We provide an explanation for the lack of attacks on Bitcoin in \Cref{sec:what_deters_attacks} and discuss other factors that deter attacks in practice in \cref{sec:appendix_what_deters_attacks}. 
\Cref{sec:StubbornNakamoto} presents the Stubborn Nakamoto protocol and proves its security. 
We provide comments on proof-of-stake protocols in \cref{sec:PoS}. \cref{sec:Discussion} concludes. 

Technical details and formal definitions are in \Cref{sec:appendix-model}. Omitted proofs can be found in \cref{sec:omitted-proofs}. \cref{sec:ex_profitable_attacks} provides examples of profit from consistency attacks.

%% file: 2_Model_and_Def.tex
\section{Model and Definitions}
\label{sec:definition}

This section describes our model and defines economic security for permissionless protocols. The model abstracts away from some technical details, which are given in \cref{sec:appendix-model}. 

\subsection{Background: Distributed ledgers and consensus}

A consensus protocol coordinates a collection of parties to collectively maintain a single write-only ledger. We interchangeably refer to the parties who maintain the ledger as \emph{miners} or \emph{nodes}. Each node  $i\in\nodes$ keeps a local copy of the ledger.  Denote the local ledger of node $i$ at time $t$ by $\ledger_i^t$. 
With a slight abuse of notation, we write $\ledger \subseteq \ledger'$ if the ledger $\ledger'$ can be generated by appending data to $\ledger$.\todo{do we ever use this?}
We refer to a generic piece of data in the ledger as a transaction $\tx$, and assume that users continuously generate unforgeable\footnote{Each transaction is cryptographically signed, ensuring that its owner authorizes it. The analysis allows for a ledger that holds arbitrary data records.} transactions and broadcast them to all nodes. Users may issue conflicting transactions, and the ledger needs to resolve such conflicts.\footnote{A user may issue a payment twice by mistake, or a malicious user may create conflicting transactions. The ledger must include only one of the conflicting transactions. Alternatively, the ledger can be viewed as a strict ordering of all finalized transactions, where a transaction is deemed invalid (and should be ignored) if it conflicts with earlier transactions.}

We say that node $i$ finalizes a transaction $\tx$ if $i$ adds the transaction $\tx$ to its ledger (updating balances accordingly). 
The ledger is append-only, meaning that data can only be added (never deleted). That is, $\ledger_i^t \subseteq \ledger_i^{t'}$ for any   $i\in\nodes$ and  $t'>t$.

Each node $i\in\nodes$ can observe only its local history, which we call the node's \emph{view}. In particular, a node may not know what other nodes observe. 
Depending on its view, the node may wait to receive messages, send messages, run computations, or finalize transactions (appending them to the ledger).
In particular, a node can hold received transactions as ``pending'' before finalizing them. 

A distributed protocol specifies each node's actions given its view. Nodes are said to be \emph{honest} if they follow the protocol and are said to be \emph{faulty} or \emph{corrupt} otherwise. Correctness\footnote{These properties are agnostic about the content of the ledger. Some applications impose additional \emph{validity} properties that ensure that the content of the ledger is valid (e.g., the ledger can include only transactions that are properly signed).} of the consensus protocol is captured by the following\footnote{These definitions are simplified to focus on the core elements. Protocols that rely on randomness (e.g., hash functions) may inevitably fail with some positive probability. The technical definition addresses this issue by requiring the probability of failure to be appropriately small; see \cref{sec:appendix-model} for details. In addition, we implicitly assume that the protocol has sufficient capacity to include all transactions.}
properties:

\begin{itemize}
\item 
{\bf Consistency:} There are no disagreements between finalized ledgers of honest nodes. That is, for any honest nodes $i$ and $j$ (where $i$ and $j$ may be the same or different) and times $t,t'$, either $\ledger_i^t \subseteq \ledger_j^{t'}$
or $\ledger_i^t \supseteq \ledger_j^{t'}$. 
\item 
{\bf Liveness:} There exists a confirmation time $T$ such that if an honest node $i$ receives transaction $\tx$ at time $t$, then any honest node $j$ finalizes $\tx$ by time $t+T$. 
\end{itemize}

Informally, consistency ensures agreement among all the nodes that maintain local copies of the ledger, and liveness ensures that the ledger keeps updating. 
When both\footnote{While a consensus protocol is useful only if it satisfies both properties, it is useful to separate the two properties to distinguish different possible failures. Liveness is violated if the system stops processing records (e.g., a payment system that stops processing transactions because it has lost communication). Consistency is violated if two nodes disagree on a finalized record, then (e.g., a website gives a user confirmation that a payment has been processed although the payment failed).}
of these properties hold, it appears to protocol users as if a single record keeper maintains a continuously updated ledger.


\subsection{Free entry and permissionless protocols}

Our interest is in protocols that provide an economically valuable property: nodes are exchangeable and may freely enter and exit. The following definition enables free entry.%
\footnote{Running a node may require computational infrastructure, and node operators may need to pay the required costs to obtain it \citep{garratt2023fixed,prat2021equilibrium}. The definition requires no entry restrictions within the protocol.}

\begin{definition}
We say the protocol is \textbf{permissionless} if a node can join or leave the protocol at any time. 
\end{definition}

In particular, nodes do not require permission from any authority, centralized or distributed. 
Bitcoin's Nakamoto protocol \citep{satoshi-bitcoin} is permissionless, allowing new nodes to join at any time.\footnote{Moreover, nodes are anonymous to the Nakamoto protocol, and a newly joined node is treated identically to any other node.} 
By contrast, classic consensus protocols (e.g., \citealt{pease1980reaching}, \citealt{dolev1983authenticated}, \citealt{castro1999practical}) are permissioned and assume that the set of nodes is known and fixed.

When analyzing a run of the protocol from time $t$ to time $t'$, we say that a node is \emph{online} if it joined before time $t$ and remains active through time $t'$. We say a node is a \emph{late joiner} if it joined at some time $\tau>t$. Note that consistency and liveness of the protocol must hold for both online nodes and late joiners. 

\subsection{Communication assumptions}

We assume synchronous communication with dynamic joins and leaves \citep{pass2017sleepy}, unless explicitly stated otherwise. 
The same communication assumption is used in previous analyses of Nakamoto's protocol~\citep{garay2015bitcoin,pass2017analysis,shitextbook}.
We denote the maximum message delay by $\Delta$. In other words, if an honest node $i$ sends a message to an honest node $j$ at time $t$, the message is guaranteed to be delivered at or before time\footnote{The model allows for longer message delays if either the sender or the receiver is faulty.} $t + \Delta$.

Nodes dynamically join and leave. If a node joins at some time $t$, then in each  $t' \geq t$ it will have received all messages sent by honest nodes in round $t' - \Delta$ or earlier. Further, when a node joins, it may receive arbitrary messages from adversarial nodes. Note that late joiners cannot observe the timing of any messages sent before they joined.\footnote{For example, suppose that an adversary attempts to broadcast a message at time $t$ and falsely claims that this message was broadcast to all nodes at time $\tau< t - \Delta$. An online node can detect that the message was not broadcast at time $\tau$ because it did not receive it by time $\tau+\Delta$. However, a late-joining node cannot detect whether the message was sent at time $t$ or earlier.}

\subsection{Definition of economic security}\todo{We use "merchant" as the victim.}

This section motivates and defines economic security for permissionless ledgers. Our goal is to provide users with meaningful guarantees, even if they are unaware of the protocol's technical specifics. 



To motivate our definition, consider a merchant attempting to receive payment from a buyer. A dishonest buyer may want to deceive the merchant into believing that payment was made without transferring funds to the merchant. Fearing the buyer may be dishonest, the merchant might be reluctant to accept paper bills that might be counterfeit or a paper check that may not be backed by sufficient funds. To reduce the risk of fraud, the merchant may employ some security measures. For example, the merchant may check security devices on the bills or call the bank to verify that a check will be honored.

In a centralized digital payment system, a trusted recordkeeper (bank or payment processor) maintains a ledger of balances and transactions. 
A dishonest buyer may attempt to mislead the merchant into believing that payment was made when, in fact, the payment is not acknowledged by the trusted recordkeeper (and no funds are transferred to the merchant). 
To mitigate such risks, the merchant employs a point-of-sale terminal (computer) that verifies the payment. 
The merchant cannot be deceived if any transaction verified by the merchant's computer is recognized and approved by the trusted recordkeeper.

In an open distributed ledger, there is no trusted recordkeeper.
Instead, a network of computer servers attempts to solve the canonical problem of distributed systems: collectively maintaining a single ledger.
Each node holds a local copy of the ledger, and honest nodes update their local ledger according to the protocol's directions. 
If consistency holds, the many local copies held by different honest nodes agree. 
Consistency allows us to refer to \emph{the} consensus ledger, although no canonical copy exists.\footnote{One analogy is that one may refer to ``the Bible'' even though there is no canonical copy of the Biblical text.} 

Security for the merchant corresponds to the consistency of the protocol.
Consider the merchant's point-of-sale terminal as a node in the distributed protocol. 
The merchant's node attempts to determine whether the transaction is included in the consensus ledger.
When consistency holds,\footnote{The consistency property is also appropriate for centralized payment systems, in which the merchant attempts to maintain consistency between the point-of-sale terminal and the trusted record keeper.} %
a merchant's terminal that follows the protocol and determines that a transaction is finalized guarantees that the transaction is included in the consensus ledger (that is, it will be included in the local copy of the ledger of any honest node). In other words, the protocol's consistency guarantees the merchant cannot be deceived: any transaction finalized by the merchant must be accepted by any honest node. 

Conversely, if an attacker deceives a merchant who follows the protocol, then consistency is violated.



We combine the distributed systems model with an economic model to define economic security. 
Classic results in the distributed systems literature prove the consistency and liveness of protocols given a bound on the number of faulty nodes. For example, classic consensus protocols such as Paxos \citep{lamport2001paxos} and PBFT \citep{castro1999practical} guarantee consistency and liveness when more than 2/3 of the nodes are honest.\footnote{These classic consensus protocols are permissioned, that is, designed for settings where a central designer authorizes each node (for example, a firm operating multiple data centers). In such settings, faults may occur due to hardware outages or software errors, and it is natural to assume that faulty behavior of many nodes is rare.} However, consistency may be violated if a majority of nodes are faulty. 

Permissionless protocols require a different approach. Under permissionless protocols, the number of faulty nodes is endogenous. An attacker can control many nodes that join the protocol or create financial incentives to convince existing node operators to deviate from the protocol. To give a meaningful quantification of the challenges the attacker must overcome in order to violate the protocol's consistency, we measure the attacker's cost of violating consistency. If the attacker cannot violate the protocol's consistency, we say the protocol has infinite economic security. If the attacker can violate consistency, quantifying the protocol's economic security requires an economic model specifying the attacker's possible actions and the cost of each possible action. Thus, our definition requires a specification of a distributed systems model and an economic model. 

\medskip


\todo{can the attacker select equilibrium? What kind of implementation?}

\begin{definition}
Given a distributed systems model and an economic cost model, the economic security of a distributed ledger is the minimal attacker’s cost required to violate consistency. 
\end{definition}

\jacob{Say: other things can break, but there are simple solutions for everything except consistency. Give the big stone example, say Bitcoin was the first to solve consistency. }


Our definition takes a similar approach to \cite{budish2022economic} in that we focus on the attacker's cost, which can be quantified within the model. By contrast, the benefits to an attacker may depend on many aspects that cannot be observed within the model.\footnote{For example, the attacker may have financial interests stemming from bets made outside the system.} (\cref{sec:ex_profitable_attacks} provides examples of the potential profitability of consistency violations.) 

\jacob{Note that it's not equilibrium. Resilient to 1/3 or 1/2 means that deviations of small players do not matter; we need an attack by a large miner.

Footnote: but equ can matter if miners want to be lazy or maximize their rewards. The resulting behavior is close enough to honest for our purposes. 
}

Knowledge of the protocol's economic security allows protocol users to adjust their behavior to ensure that an attacker cannot profitably deceive them (given reasonable assumptions on the attacker's benefit). For example, a merchant may only accept payments of a lower value than the protocol's economic security, making deception unprofitable for the attacker.

%% file: 3_Bitcoin_Analysis.tex
\section{Incentives in Permissionless Protocols: Analysis of Bitcoin's Nakamoto Protocol}
\label{sec:Bitcoin_incentives}


\jacob{THis section shows the importance of tracking KNOWLEDGE}

We formally demonstrate challenges in securing permissionless protocols by applying our definition of economic security to Bitcoin's Nakamoto protocol. Our analysis focuses on two distinct economic environments, chosen to highlight the inherent incentive deficiencies. First, the protocol lacks a mechanism to penalize attackers. \cref{sec:bitcoin-rental} describes how the protocol can be attacked, and \cref{sec:explaining_bitcoin_faults} details the design limitations that prevent it from detecting and punishing attacks. Second, even if an attack prompts a response from outside the protocol (e.g., an exchange rate collapse), this response negatively impacts all miners regardless of their involvement in the attack. Section \ref{sec:bitcoin_bribe} shows that if miners are small, the tragedy of the commons can arise. 

The results suggest that Bitcoin's security does not stem from miners' monetary incentives. We provide an alternative explanation for the security of Bitcoin in Section \ref{sec:what_deters_attacks}. Other considerations are discussed in \cref{sec:appendix_what_deters_attacks}.

\subsection{Background: Bitcoin's Nakamoto protocol }
For completeness, we provide a brief description of the Nakamoto protocol. 
Some technical details are deferred to \Cref{sec:appendix-model}.

\paragraph{The Blockchain data format}
The ledger is held in the format of a \emph{blockchain}, which grows over time as new blocks are appended.
A blockchain is an ordered sequence of blocks $(\B_0,\B_1,\dots,\B_n)$, 
where $\B_0$ is the protocol's genesis block. Each block $\B_\ell$ contains transaction data and a reference to the preceding block $\B_{\ell-1}$. A block $\B_n$ uniquely identifies a chain of preceding blocks, and we write $\chain(\B_n)=(\B_0,\B_1,\dots,\B_n)$. We say that block $\B_{n-k+1}$ is $k$-deep in $\chain(\B_n)$.
Let $h(\B_\ell)=\ell$ denote the height of the block, defined as the number of blocks in $\chain(\B_\ell)$ minus one. 

The \emph{view} of a node $i \in \nodes$ encompasses all the information available to node $i$, including its local history and any blocks it has received. A node may receive multiple blocks of the same height, which cannot belong to the same chain. 


\paragraph{Brief description of the Nakamoto protocol}

Each node (miner) organizes valid pending transactions into a suggested block. The protocol uses a hash function to select a node's block randomly. Each node calculates the hash of its block,\footnote{Following the literature~\citep{garay2015bitcoin,pass2017analysis,shitextbook}, \elaine{cite more here} we model the hash function as a random oracle that always returns a random integer between $[0, 1]$ on a fresh input. We may assume that the execution proceeds in (possibly infinitesimally small) time increments, such that a unit of mining power can invoke the hash function once per increment. A miner can calculate a new hash for the same block by changing a nonce field. This field serves the sole purpose of allowing multiple inputs to the hash function for the same block.} and the block is valid if the hash is lower than the difficulty threshold $D$.\footnote{The difficulty threshold is periodically adjusted.}
A node that finds a block with a valid hash is said to have mined a block. 
A node's mining power is the number of hash computations it can attempt per unit time. 
Nodes can write a special transaction into their block that awards them a block reward $p_B$.\footnote{A node that mines a block can include a special transaction that awards the node with a fixed number of newly minted coins and transaction fees from transactions included in its block. We collectively refer to these as the block reward and use $p_b$ to denote its USD value. The node can claim the block reward only if the block is included in the consensus chain.}


The protocol specifies the genesis block $\B_0$ that all chains must extend. If a node mines or receives a block, it broadcasts the block to all other nodes. The protocol directs nodes to mine blocks that extend the {\it longest chain} in their view. Ties are broken arbitrarily. 

The Nakamoto protocol itself does not specify when transactions are finalized. In common usage,\footnote{The parameter $k$ is commonly called the ``confirmation depth''. A common value for $k$ is $6$ blocks. For example, see \href{https://coinmarketcap.com/academy/article/how-long-does-a-bitcoin-transaction-take}{https://coinmarketcap.com/academy/article/how-long-does-a-bitcoin-transaction-take}.} finality is parameterized by a required block depth $k$, where a transaction is considered finalized by node $i$ if the transaction is within a block at least $k$-deep in the longest chain in $i$'s view. That is, if $\chain(\B_n)=(\B_0,\B_1,\dots,\B_n)$ is (one of) the longest chains in the view of node $i$, then $i$'s local copy of the ledger is $\ledger_i=(\B_0,\B_1,\dots,\B_{n-k+1})$.




\subsection{Economic security of Bitcoin under the rental model}
\label{sec:bitcoin-rental}
We consider a stylized economic environment where miners can freely enter and exit. Several papers have explored a similar model and argued that an attacker can violate the consistency of Nakamoto at a low cost (\citealt{budish2022economic}, \citealt{Tabarrok2019}, \citealt{auer2019beyond}, \citealt{moroz2020double}, \citealt{gans2024zero}). 
We provide a similar result in our framework, showing that protocol users cannot defend against these attacks and clarifying why the protocol's design enables such attacks. 

In the rental model, miners choose their computational power and may flexibly adjust it (for example, by renting capacity from a cloud computing service). Computing costs are linear: each miner pays $c$ per attempt to compute a hash. Each hash provides the miner a $1/D$ chance of mining a block. Each block includes a reward of $p_\B$. Miners are purely profit-motivated. The attacker has the same capabilities and costs as other miners. 

Consider an attacker who wishes to deceive a merchant into accepting an invalid transaction. That is, the merchant finalizes the transaction, but the transaction is not included in the consensus chain at some later time (e.g., when the merchant seeks to spend the funds). The merchant uses an observer node\footnote{An observer node is identical to a mining node with zero hash power.} to determine whether the transaction was finalized. Thus, to deceive the merchant, the attacker must violate the consistency of the protocol. 

The following theorem states that Nakamoto's economic security in this economic environment is zero, as the attacker can violate consistency at zero cost. Moreover, this holds even if the merchant uses additional security measures, such as requiring additional verifications before finalizing transactions or monitoring network activity.

\begin{prop}
\label{prop:NakaRentalAttack}
The economic security of Bitcoin's Nakamoto protocol in the rental model is zero; that is, the attacker can violate consistency and deceive a merchant at zero net cost. This holds even if the merchant perfectly monitors all network activity and requires additional confirmations for finality. 
\end{prop}

The attacker deceives the merchant by creating a conflicting ledger: a blockchain in which the transaction crediting the merchant is absent. This is possible because of the permissionless nature of the protocol, as the attacker can construct such a blockchain by operating a set of miners and providing them with malicious inputs.\footnote{The attacker's miners can even run the same code as honest miners. However, the attacker directs its miners to operate in a separate environment where they never receive the merchant's transaction or any blocks from a chain that includes it.} The attacker waits for the merchant to finalize the transaction before initiating the attack, rendering any monitoring by the merchant ineffective. The critical phase of the attack involves persuading honest nodes to adopt the conflicting ledger. This is achievable because of the longest chain rule: if the attacker’s blockchain becomes longer than the honest chain, all honest nodes will adopt the conflicting chain, thereby violating consistency.


The proof uses a similar approach as \cite{budish2022economic}, \cite{auer2019beyond}, and \cite{gans2024zero}, and is included in \cref{sec:omitted-proofs} for completeness.%
\footnote{
\cite{budish2022economic} and \cite{auer2019beyond} show that Nakamoto has positive economic security if the attacker cannot recoup the mining rewards or if the attacker has higher mining costs. 
\cite{gans2024zero} show that the net cost of the attack can be negative because the attacker can affect the block reward. 
We discuss such factors in Section \ref{sec:appendix_what_deters_attacks}. 
} 
The key argument in the proof is that the protocol fails to recognize that an attack has occurred and treats blocks mined by the attacker as if they were mined by honest miners. 
In particular, once the attacker's chain is adopted, the attacker becomes the ``honest miner'' who created the agreed ledger and is paid the mining rewards. Since honest miners do not lose from mining, neither does the attacker. 

Observe that the result holds regardless of the amount of mining rewards in the protocol. This is because mining rewards are given equally to honest and attacking miners. However, deterrence of attacks requires that an attacker's payoff is lower than an honest miner's payoff.

\subsection{Why Nakamoto fails to detect attacks}
\label{sec:explaining_bitcoin_faults}
To understand why the Nakamoto protocol permits such attacks, it is useful to examine its design and the rationale behind its longest-chain rule.

If all nodes are honest and messages are communicated instantly, exactly one block is mined at each height, and nodes achieve consensus on this unique chain. But if messages are communicated with some lag, it is possible for two honest nodes to ``simultaneously'' mine conflicting blocks of the same height.\footnote{For example, assume that it takes $10$ seconds for a message from node $i$ to reach node $j$. Suppose that at time $t$, all miners attempt to extend block $\B_{n-1}$, and that node $i$ mines a block $\B_{n}$ at time $t$ and communicates it immediately to all other nodes. At time $t+5$ honest node $j$ is yet unaware of the block $\B_{n}$ mined by $i$, and can mine block $\B'_{n}$ (that conflicts with block $\B_{n}$). At time $t+20$, the local view of any node includes two conflicting blocks $\B_{n},\B'_{n}$, and it is unclear which block should be considered part of the consensus chain.}

To address this possibility, the Nakamoto protocol uses the longest-chain rule to reach consensus on a unique chain. 
If conflicting blocks of the same height exist, the next mined block extends and selects one unique chain. 
Although there is a possibility that both blocks are extended simultaneously, the probability of continuing conflict becomes small when network lags are sufficiently short (relative to inter-block time). The probability that multiple honest miners mine conflicting blocks ``simultaneously'' is small, and the probability that two conflicting chains $(\B_0,\dots,\B_{n-1},\B_{n},\dots, \B_{n+k} )$ and $(\B_0,\dots,\B_{n-1},\B'_{n},\dots, \B'_{n+k} )$ are both mined by honest nodes is exponentially small in $k$. This observation is used to show that the Nakamoto protocol satisfies\footnote{See \cref{sec:appendix-model} and \cite{garay2015bitcoin,pass2017analysis,shitextbook}, and \cite{guo2022bitcoin}.
} consistency and liveness if (i) communication is synchronous: network lags are bounded and sufficiently short, and (ii) a sufficient fraction of the mining power is controlled by honest nodes.


Both assumptions are necessary: the Nakamoto protocol can lose consistency if network lags are too long or the adversary controls enough computational power. To see why synchrony is necessary, consider a network partition. That is, suppose that some miners are in North America (NA) and some are in South America (SA) and that a network partition prevents any communication between NA and SA for a period of time.
The protocol directs each group of miners to continue mining and extend the longest chain in their view.\footnote{Partitioned nodes in SA can observe that they do not receive any messages from NA nodes but cannot determine whether NA nodes have stopped running the protocol or whether there is a communication failure. 
The protocol directs SA nodes to continue operating as if all NA nodes stopped.} 
Two conflicting chains grow, one by NA nodes and one by SA nodes. When the network partition is resolved, both NA nodes and SA nodes become aware of both chains. At this point, the protocol reconciles the two conflicting chains by directing all nodes to adopt the longest chain. If the partition lasts long enough, this reconciliation can violate consistency: NA and SA finalize conflicting blocks that are $k$-deep, and the group with the shorter chain discards some finalized blocks.

An attacker with sufficient computational power essentially exploits the longest-chain rule by ``pretending'' to be partitioned. If the attacker can mine a conflicting chain that is longer than the honest chain, the protocol directs all miners to adopt the new longest chain, as in the reconciliation. 



In practice, the community will be aware that the protocol was attacked. Under synchronous communication, which is necessary for Nakamoto \citep{pass2017rethinking}, honest nodes can detect that the attacker's blocks could not have been mined by honest nodes because they receive these blocks after a long delay. The protocol cannot detect whether this delay results from the attacker's malicious behavior or from honest nodes suffering a network partition. However, it is very unlikely that a significant fraction of the computing power in the network is partitioned.\footnote{Moreover, if such a significant partition occurs, it will likely be known to the nodes and protocol users.} A common assumption in the literature is that an attack on the protocol will trigger a collapse of the exchange rate of the protocol's coin, which implicitly assumes that the community can detect attacks.


\subsection{Bribery model and the tragedy of the commons}
\label{sec:bitcoin_bribe}
\todo{say: blocking free entry doesn't necessarily give security}


Consider now an economic environment in which the set of miners is fixed, and the attacker can obtain mining power only from existing miners. Even though the Nakamoto protocol is permissionless, external factors may restrict miner entry. For example, miners may employ specialized mining equipment or have access to other restricted resources (e.g., cheap electricity and computing facilities). This stylized economic environment can be considered as an extreme fixed cost\footnote{\cite{garratt2023fixed} analyze a security model with fixed costs and \cite{prat2021equilibrium} analyze miner entry with fixed costs.} of mining, which makes entry prohibitively expensive. 

To focus on the role of economic incentives, we assume that miners are rational and profit-driven. The attacker can affect the behavior of miners by committing to an incentive contract, which we refer to as a bribery contract \citep{bonneau2016buy,minerbribe02,minerbribe00,minerbribe03}.
We refer to this economic environment as the \emph{bribery model}. 

The bribery model allows miners to suffer losses due to a community response to an attack. When there is no free entry of miners, it is possible that $p_\B>D\cdot c$ and miners earn positive profits from mining, and the expectation of future profits increases the value of mining equipment. If the community responds to an attack by decreasing usage, the value of future mining rewards decreases, and mining equipment loses value. Miners may also incur additional losses, and we use $\Psi_A$ to capture all losses miners suffer if the system is attacked.\footnote{\cite{garratt2023fixed} endogenously determine the miners' loss and evaluate the security implications for an attacker who owns all the mining equipment used in the attack.} 

A community response may carry additional implications for the attacker. We consider these separately below and in \cref{sec:what_deters_attacks}. In the bribery model, a successful attack allows the attacker to collect the standard value of block rewards $p_\B$ despite the community response. 

The bribery model has a continuum of miners, each controlling an infinitesimal unit of computational power.\footnote{
Contemporaneously and independently of this work, \cite{newman2023decentralization} analyzes a model with finitely many miners and shows that small miners are unlikely to be pivotal to an attack's success.}
Let  $\psi_A$ denote the normalized harm per infinitesimal miner.  Miners will participate in the attack if it is in their selfish best interest, given the payments they stand to receive from the protocol, the bribes from the miner, and the harm they may suffer from an attack. 

A bribery contract induces a game among the miners. For simplicity, we assume miners play a simultaneous move game where all miners observe the bribery contract and simultaneously choose whether to participate in the attack. The attacker succeeds in attacking the protocol if the attacker violates the protocol's consistency under any equilibrium of the induced game.

The following theorem shows that even if miners are substantially harmed by an attack, it is cheap for the attacker to incentivize individual small miners to participate in an attack.

\begin{prop}
\label{prop:NakaBriberyCost}
In the bribery model, the economic security of Bitcoin's Nakamoto protocol is arbitrarily close to zero; that is, the attacker can violate consistency and deceive a merchant at a net cost arbitrarily close to zero. 
\end{prop}

\begin{proof}[Prof of \cref{prop:NakaBriberyCost}]
As in the proof of Proposition~\ref{prop:NakaRentalAttack} in \cref{sec:appendix-model}, the attacker waits until the merchant finalizes the transaction in some block $\B_n$ at some time $t$. Only then, the attacker publishes a bribery contract that pays miners who mine blocks $\B'_{n},\dots, \B'_{n+k-1}$ that form the attack chain $(\B_0,\dots,\B_{n-1}, \B'_{n},\dots, \B'_{n+k-1} )$. The contract commits to pay $\tilde{p}_b$ for each mined block.\footnote{Such a commitment can be implemented by a smart contract running on a different blockchain \citep{minerbribe02}. To receive payment, the miner submits the block to the smart contract, which can verify that the block extends the attack chain.} The block reward from each of the blocks $\B'_{n},\dots, \B'_{n+k-1}$ is directed to an address controlled by the attacker, enabling the attacker to collect them if the attack succeeds. 

\begin{table}[!ht]
\centering
\begin{tabular}{l|c|c|}
                  & \begin{tabular}[c]{@{}l@{}}Attack \\ Succeeds\end{tabular} & \begin{tabular}[c]{@{}l@{}}Attack \\ Fails\end{tabular} \\ \hline
\rule{0pt}{3ex}%
Mine attack chain &    $x\cdot\left(\tilde{p}_\B/D-c\right)-\psi_A$                                       &    $x\cdot\left(\tilde{p}_\B/D-c\right)$                                                      \\ \hline
\rule{0pt}{3ex}%
Mine honest chain &    $\phantom{\;\;\left(\tilde{p}/D\right)}-x \cdot c-\psi_A$                             &    $x\cdot\left(p_\B/D-c\right)$                                                   \\ \hline
\rule{0pt}{3ex}%
Do not mine       &    $\phantom{x\cdot\left(\tilde{p}_\B/D-c\right)}-\psi_A$                             &    $0$      \\ \hline
\end{tabular}
\caption{Miner payoffs under the bribery contract.}
\label{table:minerBribePayoff}
\end{table}

This bribery contract induces a game among miners. Each miner may choose to mine the honest chain, mine the attack chain, or stop mining.\footnote{For simplicity, we assume miners do not split their mining power.} If sufficiently many miners mine the honest chain and refuse to mine the attack chain, the attack fails. If sufficiently many miners mine the attack chain, the attack succeeds. The bribery contract guarantees to pay miners who mine a block in the attack chain an attack block reward of $\tilde{p}_\B$ regardless of whether the attack succeeds or fails. Miners who mine a block in the honest chain collect an honest block reward of $p_\B$ only if the attack fails. If the attack succeeds, all miners suffer the cost of the attack $\psi_A$ (recall that miners are anonymous). Let $x$ denote the number of hashes a miner can compute during the attack. 
The potential payoffs of a miner are summarized in Table \ref{table:minerBribePayoff}.  

By assumption, miners are small and not pivotal to the attack's success. Because honest miners must be willing to mine the honest chain when there is no attack, we have that $p_\B/D-c \geq 0 $. 
Thus, if the attacker pays $\tilde{p}_\B>p_\B$, it is a dominant strategy for each individual miner to participate in the attack. The attacker's net cost per block in the attack chain is $\tilde{p}_\B-p_\B\approx 0$, and the net cost of the entire attack is arbitrarily close to zero. 
\end{proof}

Notably, the attacker can induce miners to participate in the attack without compensating miners for the total harm of the attack, $\Psi_A$, or even for their individual losses, $\psi_A$. The game induced by the bribery contract creates a tragedy of the commons for miners. Even if $\Psi_A$ is substantial, it does not affect an individual miner's incentive to deter the attack. Miners are anonymous, and the attack harms both miners who participate in the attack and honest miners alike. If an individual miner cannot affect whether the attack succeeds or fails, the harm does not incentivize individual miners to deter attacks. 
In that respect, \cref{prop:NakaBriberyCost} gives a foundation for the EAAC (expensive to attack in the absence of collapse) property of \cite{budish2024economic}.

The community response may limit the attacker's ability to recoup its costs by collecting mining rewards. For example, suppose that the community responds to an attack by refusing to accept payments through the protocol. In that case, the attacker cannot collect block rewards after the attack. This response increases the attacker's net cost to $p_\B$ per block in the attack chain, but the attacker does not need to compensate miners for the losses from the attack, $\Psi_A$. 
Such a response differentiates between honest miners and attackers, creating penalties that deter attacks. However, other community responses can be more effective. We discuss these next.





\jacob{this sounds like a very general problem; it's hard to make many players pivotal. See 
Pivotal Players and the Characterization of Influence
Nabil I. Al-Najjar, Rann Smorodinsky 
https://www.sciencedirect.com/science/article/pii/S002205310092605X}

\subsection{What deters attacks in practice?}
\label{sec:what_deters_attacks}

Despite the possibility of attacks, major cryptocurrencies like Bitcoin and Ethereum\footnote{Ethereum used the Nakamoto protocol until 2022.} have remained secure. At the same time, numerous attacks on multiple cryptocurrencies that operate the Nakamoto protocol show that the Nakamoto protocol is indeed vulnerable to attacks in practice \citep{shanaev2019cryptocurrency}. 

We discuss several explanations for the absence of attacks on Bitcoin in Appendix \ref{sec:appendix_what_deters_attacks} and highlight one explanation in particular: the community's ability to correct the ledger after an attack, which we refer to as a \emph{community response}.
Recall that the protocol provides a record-keeping service: maintaining an agreed-upon ledger. If the community agrees that the ledger was corrupted, the community may override the protocol to change it. In blockchain terminology, the community can ``fork'' the protocol by directing all miners to ignore the attack chain and extend the honest chain.
Doing so reverts any effect of the attack, leaving the ledger unchanged by the attack.\footnote{Such corrections also occur in traditional record-keeping services. For example, the New York Stock Exchange canceled trades that resulted from a software glitch \citep{CNBC2024}.} Moreover, the community response imposes a significant risk to the attacker of having to pay all the attack's costs only to see it undone. 

Such community responses are not without precedent and are actively considered by system developers.\footnote{
For example, Buterim writes ``if the cryptoeconomic consensus fails, whether due to a bug or an intentional 51$\%$ attack, a vast community of many thousands of developers and many more users are watching carefully to make sure the chain recovers correctly'' \citep{Buterin2023DontOverload}.
The developer community often uses the terms social consensus \citep{Buterin2016Philosophy} or a community hard fork when referring to the protocol's reliance on external sources.
} 
A notable example is the Ethereum community's decision to fork the Ethereum ledger in order to return funds stolen from the DAO attack \citep{EthereumFoundation2016HardFork}.
Such measures, however, require the community to be willing to override the protocol and able to agree on the correct ledger.
While the Ethereum blockchain community was able to override the protocol, the Ethereum Classic blockchain rejects any community interventions \citep{EthereumClassic_nodate_CodeIsLaw}.
Anecdotally, Ethereum did not suffer any consistency attacks, whereas Ethereum Classic suffered multiple such attacks \citep{Nesbitt2019DeepChainReorg}.

However, the community response is not a silver bullet. To undo the attack, the community must reach an agreement on the correct ledger. In other words, the community needs to solve the same consensus problem that the protocol attempts to solve. Moreover, the attacker may employ a different attack to deceive both the protocol and the community. 
Without a formal model of the community’s decision-making process, it is difficult to assess the community response. 

In the next section, we analyze a model inspired by the efficacy of the community response. The protocol can be seen as incorporating the community response within the protocol.
The analysis evaluates the required capabilities and limitations of the community.

%% file: 4_StubNaka_3.tex
\section{Possibility of a Permissionless and Economically Secure Consensus Protocol}
\label{sec:StubbornNakamoto}



We now consider the broader question of protocol design: Can a permissionless\footnote{A common alternative approach is to modify the protocol’s permissionless properties. For example, Proof of Stake (PoS) protocols require nodes to register an identifier and associate it with an account. We discuss such protocols in \cref{sec:PoS}.} protocol achieve economic security? 
This section answers affirmatively. We show that a permissionless protocol can be economically secure by constructing such a protocol. This protocol is a variation of the Nakamoto protocol, and we name it \emph{Stubborn Nakamoto} protocol. We prove that it is infinitely economically secure -- that is, even an attacker with unlimited resources cannot violate consistency -- under the same communication assumptions required by Bitcoin's Nakamoto, namely, synchrony with late joiners. 
The Stubborn Nakamoto protocol preserves the permissionless properties of the Nakamoto protocol while ensuring economic security. Specifically, miners remain anonymous and can enter or exit the system without restriction.

The design approach underlying the Stubborn Nakamoto protocol is motivated by the analysis in \cref{sec:Bitcoin_incentives}. The negative results in \cref{sec:Bitcoin_incentives} demonstrate that the inability to detect attacks compromises the economic security of Bitcoin's Nakamoto protocol, whereas the effectiveness of the community response highlights the power of attack detection. Accordingly, the Stubborn Nakamoto protocol is designed to enable nodes to detect potential attacks.

The protocol can be interpreted as incorporating the community response which restores the correct ledger following an attack. A straightforward design might require nodes to communicate and reestablish the correct ledger after an attack. The Stubborn Nakamoto protocol takes a more efficient approach: if a node is assured that the community response will restore the correct ledger, it can immediately adjust its local copy to reflect that correct ledger. (For example, a node can reject an attack chain if it is confident that all honest nodes will likewise reject it.) The Stubborn Nakamoto protocol provides nodes with sufficient knowledge about the behavior of other nodes, ensuring that they maintain a consistent ledger.

This analysis in this section examines the role of the ``community'' -- that is, external inputs beyond the protocol. Some external input is unavoidable. For instance, the initialization of any protocol requires an external source (e.g., setting the genesis block in Bitcoin's Nakamoto protocol), and late joiners must rely on an external input to connect to the protocol (e.g., distinguishing Bitcon's chain from Dogecoin's chain). However, reliance on such external inputs should be minimized.\footnote{ Reliance on the community to resolve challenges is often challenging to formally model and analyze. Moreover, reliance on out-of-protocol inputs can introduce fundamental issues. First, the primary objective of the protocol is to establish consensus, so it should not depend on the community's ability to achieve consensus (conversely, if the protocol cannot establish consensus, why should the community be able to do so?). Second, reliance on external inputs may undermine the protocol's permissionless properties, potentially giving control over the system to an out-of-protocol entity.} 
This section rigorously evaluates the necessary role of the community, formalizing it as a \emph{recovery oracle}. We prove that some external dependence is necessary for recovery after attacks, and design the Stubborn Nakamoto protocol to minimize this reliance in two key ways. First, the requirements from the recovery oracle can be practically implemented. Second, the Recovery Oracle should not hold excessive control over the system. The Stubborn Nakamoto protocol is thus designed to minimize both: the recovery oracle is simple to implement, invoked only for attack recovery, and holds minimal influence over the system.




Section \ref{sec:def_SN} defines the Stubborn Nakamoto protocol, and we prove its security and liveness for online nodes in \cref{sec:SN_consistency_liveness}. \cref{sec:late_joiners} defines the protocol for late joiners and proves the corresponding guarantees. Section \ref{sec:SN_recovery} details how the Stubborn Nakamoto protocol can recover from an attack and formalizes the role of the community in the recovery. 
Section \ref{sec:oracle_necesity} examines the necessary role of the community.

\subsection{The Stubborn Nakamoto protocol}
\label{sec:def_SN}

\newcommand{\C}{\mathcal{C}}
\newcommand{\PC}{\mathcal{PC}}
\newcommand{\CC}{\mathcal{CC}}
\newcommand{\fin}{{\sf Log}}
\newcommand{\stub}{{\sf Stub}}
\newcommand{\view}{{\sf View}}
\newcommand{\h}[1]{h(#1)}
\newcommand{\conflict}{\not\sim}
\newcommand{\noconflict}{\sim}
\newcommand{\ton}{{t_{on}}}
\renewcommand{\B}{{\sf B}} 
\renewcommand{\chain}{\sf Chain}
\renewcommand{\ledger}{{\sf Log}}


In contrast to the analysis of Bitcoin's Nakamoto protocol in \cref{sec:Bitcoin_incentives}, the definition and analysis of Stubborn Nakamoto protocol require careful consideration of each node's knowledge, 
and in particular each node's knowledge of the state of other nodes. Before giving the definition of the Stubborn Nakamoto protocol, we introduce notation and definitions that represent the knowledge held by a node. 
Note that this section defines the Stubborn Nakamoto protocol for online nodes. The protocol for late joiners is defined in \cref{sec:late_joiners}. 

The Stubborn Nakamoto protocol makes the same communication assumptions as Bitcoin's Nakamoto protocol, namely synchronous communication with late joiners. We also assume ``implicit echoing'', namely, if an honest node\footnote{We assume that the implicit echoing assumption holds for all nodes, including observer nodes that do not mine. \cite{sridhar2024consensus} show the necessity of this assumption.} receives a message at time $t$, then every honest node receives the message by time $t + \Delta$.
Stubborn Nakamoto also shares the same setup assumptions as Bitcoin’s Nakamoto protocol: all nodes are assumed to have knowledge of the genesis block and the protocol's rules. The protocol is parameterized by a confirmation depth \( k \).

Stubborn Nakamoto uses the blockchain data format. 
The local view of node $i$ includes all the blocks received by $i$, as well as additional information recorded in the form of block labels. 
We write $\B\in\view_i^t$ to denote that node $i$ received block $\B$ by time $t$. Each block label is encoded as a subset of blocks in view. We describe below how a node assigns each label given its local view. Labels are never removed.\footnote{For example, we have $\C_i^{t}\subseteq \C_i^{\tau}$  for $\tau\geq t$ (and analogously for all other block labels defined below). } 

A block $\B$ \emph{conflicts} with block $\B'$, denoted $\B \conflict  \B' $, if there is no chain that contains both $\B,\B'$. That is, neither $\chain(B)\subseteq \chain(\B')$ nor $\chain(B')\subseteq \chain(\B)$. 
Given a block $\B$ and a set of blocks $\mathcal{A}$, if $\nexists \B'\in \mathcal{A}$ such that $\B \conflict \B'$ we say that $\B$ does not conflict with $\mathcal{A}$, written $\B\noconflict\mathcal{A}$. 
We write $\B\conflict\mathcal{A}$ if $\exists \B'\in \mathcal{A}$  such that $\B \conflict \B'$. 


The protocol may instruct a node to \emph{halt} if the node detects a potential conflict between its local view and the local view of another honest node. When a node halts it stops its attempts to mine any new blocks and stops certifying any new blocks (as described below). If the node has already certified some blocks, it completes the steps below to determine whether these blocks should be finalized.

\smallskip

The Stubborn Nakamoto protocol relies on the following definition, which specifies the first step a node takes toward finalizing a block. The conditions for this first step closely resemble the conditions for finalizing a block in the Nakamoto protocol. However, unlike the Nakamoto protocol, the required confirmation depth $k$ is explicitly defined within the protocol, and all nodes use the same confirmation depth $k$.


\begin{definition}
\label{def:certified}
Node $i$ \textbf{certifies} block $\B$ at time $t$, written $\B\in \C_i^t$, if at time $t$ there is a chain $(\B_0,\dots,\B,\dots,\bar{\B})$ in $\view_t^i$ such that
    \begin{enumerate}
        \item \label{def:cert1} Block $\B$ is at least $k$-deep in the chain $\chain(\bar{\B})$, i.e., $h(\bar{\B}) - h(\B) +1 \geq k$. 
        \item \label{def:cert2}  The chain $\chain(\bar{\B})$ is a longest chain in view, meaning $\h{\bar{\B}}\geq \h{\tilde{\B}}$ for any $\tilde{\B}\in {\view}_i^t$.
    \end{enumerate} 
\smallskip
Additionally, if in the view of node \( i \) at time \( t \) there is no certified block \( \B' \) that conflicts with \( \B \), then \( \B \) is also \textbf{clear-certified}, denoted \( \B \in \CC_i^t \).\footnote{If two conflicting blocks both become certified at time \( t \), neither is clear-certified.}
\end{definition}

The protocol relies on the following definition to characterize each node’s knowledge of what other nodes may have certified.

\begin{definition}
\label{def:potentially_certified}
For $\gamma\geq0$, a block $\B$ becomes \textbf{$\pmb{\gamma}$-potentially certified} by node $i$ at time $t$, written $\B\in \PC_i^t(\gamma)$, if at time $t$ there is a chain $(\B_0,\dots,\B,\dots,\bar{\B})$ in ${\view}_t^i$ such that 
    \begin{enumerate}
        \item block $\B$ is at least $k$-deep in the chain $\chain(\bar{\B})$, that is $h(\bar{\B}) - h(\B) +1 \geq k$; 
        \item the chain $\chain(\bar{\B})$ would have been a longest chain in $i$'s view at time $t-\gamma$, meaning $\h{\bar{\B}}\geq \h{\tilde{\B}}$ for any $\tilde{\B}\in {\view}_i^{t-\gamma}$.
    \end{enumerate}
\end{definition}

We can now state the definition of the Stubborn Nakamoto. 
Recall that each node $i$ maintains $\ledger_i^t$, a local copy of the ledger that consists of all finalized blocks. The local copy of the ledger $\ledger_i^t$ is required to agree with the local copy of the ledger $\ledger_j^\tau$ of any other honest node $j$. 

The Stubborn Nakamoto protocol requires the same communication assumptions as Bitcoin's Nakamoto: synchronous communication with late joiners. We make the standard ``implicit echoing'' assumption (which is also assumed for standard Nakamoto).\footnote{That is, every node (including observer nodes that do not mine) echoes every fresh message they see. This implies that if an honest node sees a message at time $t$, then every honest node will have seen it by $t + \Delta$.} 
It also shares the same setup assumption as Bitcoin's Nakamoto: all nodes know the genesis block and the protocol. Stubborn Nakamoto is parameterized by a confirmation depth $k$. 

The Stubborn Nakamoto protocol directs nodes to take two steps before finalizing a block and adding it to their local copy of the ledger. First, the block needs to be certified. Second, the node waits to see that no other node could have certified a conflicting block. If the node does not observe any potential conflicts, the node can finalize the block and add it to the ledger.

\begin{definition}[The Stubborn Nakamoto Protocol]
As in Nakamoto, the protocol directs nodes (miners) to attempt to mine a block that extends the longest chain in view. Each node $i$ takes the following steps%
\footnote{\interfootnotelinepenalty=10000
Using notation: Let $t_1$ be the earliest time such that $\B\in\C_i^{t_1}$. 
\begin{enumerate}[leftmargin=40pt]
    \item If $\B\conflict\C_i^{t_1}$, then node $i$ halts at time $t_1$. 
    \item Let $t_2=t_1+2\Delta$. If $\B\conflict  \PC_i^{t_2}(2\Delta)$, then node $i$ halts at time $t_2$. If $\B\noconflict  \PC_i^{t_2}(2\Delta)$, then $\B\in  {\fin}_i^{t_2}$.
    \item Let $t_3=t_1+4\Delta$. If $ \B\conflict  \PC_i^{t_3}(4\Delta)$, then node $i$ halts at time $t_3$.
\end{enumerate}
}
 towards finalizing blocks:
\begin{enumerate}
    \item Node $i$ certifies blocks according to definition \ref{def:certified}. If a block $\B$ is certified but is not clear-certified (that is, in the local view of node $i$ there is a block $\B'$ that is certified and conflicts with $\B$), then node $i$ halts. 
    \item If  $i$  certified block $\B$ at time $t_1$, then at time $t_2=t_1+2\Delta$ node $i$ checks if there is a block $\B'$ in $i$'s local view that conflicts with $\B$ and is  $2\Delta$-potentially certified. If yes, node $i$ halts. If $\B$ does not conflict with any $2\Delta$-potentially certified block in $i$'s local view, then node $i$ finalizes $\B$, written $\B\in {\fin}_i^{t_2}$.
    \item \label{step:stub} If  $i$  certified block $\B$ at time $t_1$, then at time $t_3=t_1+4\Delta$ node $i$ checks if there is a block $\B'$ in $i$'s local view that conflicts with $\B$ and is  $4\Delta$-potentially certified. If yes, node $i$ halts. 
\end{enumerate}
\end{definition}

The definition of the Stubborn Nakamoto protocol is conceptually simple yet carefully designed. On one hand, the protocol ensures that nodes can detect and prevent conflicts with other nodes. On the other hand, nodes must not be overly cautious, as halting too easily would hinder progress and compromise the protocol's liveness. The protocol’s design, along with the definition of $\gamma$-potentially confirmed blocks, carefully balances these competing requirements.

One interpretation of the protocol is that it incorporates the community response (discussed in Section \ref{sec:what_deters_attacks}) into the protocol itself. Because the Stubborn Nakamoto protocol is permissionless, it cannot prevent an attacker with sufficient resources from creating a conflicting ledger (as described in \cref{sec:bitcoin-rental}). 
Despite this, the protocol prevents consistency violations by allowing nodes to identify attacks. If a node can identify received blocks as attack blocks that could only have been mined by an attacker, it can simply ignore them. However, to maintain agreement, we must ensure that if node $i$ ignores a block, it can be confident that any honest node $j$ will also ignore that block. 

To ensure agreement among all honest nodes, the Stubborn Nakamoto protocol differs from the standard Nakamoto in two key ways. First, it specifies an explicit condition for finality, ensuring that an honest node can reliably predict the behavior of other honest nodes.
Second, before finalizing a block, nodes wait for a conflict-discovery period, allowing them to learn if another node receives a conflicting ledger. 
While an attacker with unlimited resources may still create a conflicting ledger, honest nodes will be able to detect the existence of multiple potential ledgers, identify the risk of disagreement, and halt if necessary. More generally, the protocol guarantees that an honest node finalizes a block only if that node is assured that no other honest node could have finalized a conflicting block. 

The Stubborn Nakamoto protocol prioritizes consistency and instructs nodes to finalize a block only if consistency is assured. Given the known impossibility of consensus without an honest majority \citep{pass2017rethinking,lewis2023permissionless, sridhar2024consensus}, any protocol must, under some attacks, sacrifice either consistency or liveness. The protocol may lose liveness under some attacks that cause nodes to detect a conflict and halt. We discuss how the protocol can regain liveness after an attack in \cref{sec:SN_recovery}.

\begin{remark}
\label{rem:ignore_conf_blocks}
The Stubborn Nakamoto protocol can be made more resilient by instructing nodes to become ``stubborn'' on any block for which they completed step \ref{step:stub}, meaning a node ignores any block that conflicts with a block that completed step \ref{step:stub} without halting. This modification allows online nodes to completely ignore the attack described in the \cref{sec:bitcoin-rental}, and it does not affect the protocol's security and liveness guarantees for online nodes (which are formally stated and proved in the following section). However, such a modification will require careful treatment of late joiners, which we discuss in \cref{sec:late_joiners} and \cref{sec:SN_recovery}. 
\end{remark}


\subsection{Consistency and liveness of Stubborn Nakamoto }
\label{sec:SN_consistency_liveness}

We turn to prove that Stubborn Nakamoto guarantees consistency and liveness for online nodes. Treatment of late joiners is in \cref{sec:late_joiners}. 
Before presenting our main theorem, we first establish formal results that characterize the knowledge an honest online node has about other honest online nodes under the Stubborn Nakamoto protocol.

\begin{lem}
\label{lem:certify_echo}
If an honest node $i$ certifies block $\B$ at time $t$, then by time $t+\Delta$ block $\B$ is $2\Delta$-potentially certified by any honest node $j$. \\
If an honest node $i$ $\gamma$-potentially certifies block $\B$ at time $t$, then by time $t+\Delta$ block $\B$ is $(\gamma+2\Delta)$-potentially certified by any honest node $j$. 
\end{lem}



\begin{proof}
Because $\C_i^t = \PC_i^t(0)$, it is sufficient to prove the second part of the lemma. 
Assume that block $\B$ is $\gamma$-potentially certified by honest node $i$ at time $t$.  That is, there is a chain $\chain(\bar{\B})=(\B_0,\dots,\B,\dots,\bar{\B})\subseteq \view_i^t$ that satisfies the conditions of definition \ref{def:potentially_certified} given the local view of node $i$ at time $t$, namely: $h(\bar{\B}) - h(\B) +1 \geq k$, and $\h{\bar{\B}}\geq \h{\tilde{\B}}$ for any $\tilde{\B}\in {\view}_i^{t-\gamma}$ . 

Consider arbitrary honest node $j$.
Because node $i$ receives all the blocks in $\chain(\bar{\B})$ by time $t$, node $j$ receives all blocks in $\chain(\bar{\B})$ by time $\tau\leq t+\Delta$. We have that $\B$ is $k$-deep in $\chain(\bar{\B})$, satisfying $h(\bar{\B}) - h(\B) +1 \geq k$. 
To see that $\h{\bar{\B}}\geq \h{\tilde{\B}}$ for any $\tilde{\B}\in {\view}_j^{\tau-\gamma-2\Delta}$, observe that 

$$ {\view}_j^{\tau-\gamma-2\Delta} \subseteq  {\view}_j^{t+\Delta -\gamma-2\Delta} = {\view}_j^{t -\gamma-\Delta} \subseteq {\view}_i^{t -\gamma} $$
where the first inclusion follows because $ \tau\leq t+\Delta$, and the last inclusion follows from communication between two honest nodes.  Thus, we have that $\B\in \PC_j^\tau(2\Delta+\gamma)\subseteq \PC_j^{t+\Delta}(2\Delta+\gamma)$. 
\end{proof}

We can now state our main result, showing that Stubborn Nakamoto is economically secure. 

\elaine{TODO: add why synchrony is needed to get infinite economic security}
\jacob{for online nodes}
\begin{theorem}[Consistency of Stubborn Nakamoto]
Under synchrony, Stubborn Nakamoto satisfies consistency even in the presence of an adversary with unbounded computational power\footnote{That is, the attacker can control arbitrarily mining resources with an unbounded hash rate. We maintain the standard assumption that the attacker cannot break cryptographic primitives.}  and any number of corrupted nodes.
\label{thm:consistency-stubborn}
\end{theorem}

\begin{proof}
Suppose that honest node $i$ finalized block $\B$ at time $t_2$. Let $t_1=t_2-2\Delta$  denote the time when node $i$ certified block $\B$. 
We need to show that if some honest node $j$ finalized a block $\B'$ 
then it must be that $\B'\noconflict\B$. It is immediate from definition \ref{def:certified} that node $i$ never clear-certifies a block $\B'\conflict \B$, 
and therefore node $i$ cannot finalize a block $\B'$ that conflicts with $\B$. 

For the sake of contradiction, consider an honest node $j\neq i$ that finalized a block $\B'\conflict \B$ 
at time $t_2'$, and let $t_1'=t_2'-2\Delta$  denote the time node $j$  certified block $\B'$. 
By \cref{lem:certify_echo}, node $j$ must have $2\Delta$-potentially certified block $\B$ by time $t_1+\Delta$. Because node $j$ finalized block $\B'\conflict\B$, we must have that $t_1'+2\Delta=t_2'< t_1+\Delta$. Symmetrically, by \cref{lem:certify_echo} we have that node $i$ must have $2\Delta$-potentially certified block $\B'$ by time $t_1'+\Delta$. Because node $i$ finalized block $\B\conflict\B'$, we must have that $t_1+2\Delta=t_2< t_1'+\Delta$. But this implies that $t_1<t_1'-\Delta$ and $t_1'<t_1-\Delta$, a contradiction.

\end{proof}

The consistency guarantee provided by \cref{thm:consistency-stubborn} is notable for two key reasons.
First, it demonstrates that the Stubborn Nakamoto protocol is maximally economically secure, as even an attacker with unlimited resources cannot violate consistency and deceive the protocol’s users.
Second, the protocol's security does not rely on mining rewards or other monetary incentives.\footnote{Mining rewards play a role in deterring attackers from compromising the protocol’s liveness; see \cref{sec:SN_recovery}.}

\medskip
Of course, a protocol is only useful if it also satisfies liveness. 
To see why Stubborn Nakamoto also satisfies liveness, observe that Stubborn Nakamoto differs from Nakamoto only when, in the local view of some honest node $i$, there are two conflicting blocks $\B' \conflict \B$ that are both $\gamma$-potentially certified for $\gamma\leq 4\Delta$. 
In other words, nodes will act differently under the two protocols only in realizations that are close to the Nakamoto protocol's consistency violations. Under approximately the same conditions that guarantee consistency and liveness of the Nakamoto protocol, such realizations do not arise, and Stubborn Nakamoto satisfies consistency and liveness.

\cref{thm:liveness-stubborn} builds on these ideas to formally establish that Stubborn Nakamoto satisfies consistency and liveness under a sufficient honest majority assumption, the same assumption required for Nakamoto to satisfy consistency and liveness. The exact technical statement of the network-aware honest majority assumption and the proof can be found in \Cref{sec:appendix-model}. 

\jacob{for online nodes}
\begin{theorem}[Liveness of Stubborn Nakamoto]
\label{thm:liveness-stubborn}
Under synchrony, Stubborn Nakamoto satisfies consistency and liveness if the network-aware honest majority assumption holds. 
\end{theorem}


Roughly speaking, the theorem states both Nakamoto and Stubborn Nakamoto require the same honest majority assumption to guarantee both consistency and liveness.
The protocols differ when this assumption is not satisfied.\footnote{Some trade-offs in protocol design are necessary, as no protocol can guarantee both consistency and liveness under corrupt majority \citep{pass2017rethinking,lewis2023permissionless,sridhar2024consensus}.} Under the Nakamoto protocol, an attack that controls sufficiently many nodes can cause consistency violations.
In contrast, the Stubborn Nakamoto protocol prevents consistency violations by allowing nodes to detect when consistency might be at risk. Rather than allowing transactions to be added to the ledger only to be later reverted, Stubborn Nakamoto instructs nodes to halt and wait for a safe recovery. Effectively, the Stubborn Nakamoto protocol transforms Nakamoto's consistency violations into liveness violations.
This behavior is particularly desirable in financial systems, where transaction integrity is paramount.

Note that the version of Stubborn Nakamoto outlined in \cref{rem:ignore_conf_blocks} can completely avoid some of Nakamoto's consistency violations without suffering a liveness violation. In particular, this version of the protocol maintains both consistency and liveness under the attack outlined in \cref{sec:bitcoin-rental}. 

The proof of \Cref{thm:liveness-stubborn} uses the intermediate stochastic bounds used to prove the consistency and liveness of Nakamoto. While the scenarios under which liveness of Stubborn Nakamoto is violated are very close to those under which consistency of Nakamoto is violated, it is possible to have a liveness violation of Stubborn Nakamoto without a corresponding consistency violation of Nakamoto.  For this reason, the proof of Theorem~\ref{thm:liveness-stubborn} requires subtle changes to the analysis of Nakamoto consensus of \cite{pass2017analysis} and \cite{shitextbook}.  We state the technical details, required lemmas, and the proof of \Cref{thm:liveness-stubborn} in \Cref{sec:appendix-model}.



\subsection{Late joiners}
\label{sec:late_joiners}

This section establishes guarantees for late joiners under Stubborn Nakamoto. 
A late joiner is a node that joins the system at some time $t$, after the protocol has already been running for some time and some blocks may already have been finalized. We adopt the standard communication assumption that by time $t+\Delta$ the late joiner receives all messages sent by honest nodes by time $t$ or earlier \citep{pass2017sleepy}.\footnote{
The same communication assumption is also used by the standard Nakamoto protocol, which requires late-joining nodes to receive and verify the entire chain of previous blocks, linking back to the genesis block. The delivery of past messages is essential in any protocol to ensure that newly joining nodes can correctly establish and continue the ledger.} However, a late joiner node cannot distinguish the time at which earlier messages were sent.\footnote{That is, if $\tau_1,\tau_2<t$, a late joiner cannot tell whether a message was sent at time $\tau_1$ or at time $\tau_2$. This is in contrast to online nodes, which are able to observe when any message is sent by an honest node up to a $\Delta$ time lag. In particular, late joiners cannot recreate the finalization steps taken by online nodes under Stubborn Nakamoto, as they cannot determine which blocks were received within $4\Delta$ of a block's certification.}
Additionally, a late joiner may also receive arbitrarily many messages from adversarial nodes.

Allowing late joiners is crucial for the free entry of nodes. Without this capability, all nodes would be required to join at the protocol's inception and remain continuously online. 
However, ensuring consistency for late joiners presents significant challenges. For example, the well-known Dolev-Strong protocol guarantees consistency and liveness for online nodes but fails to provide these guarantees for late joiners (see, e.g., \cite{pass2017rethinking}).  

The following definition formalizes the process by which a late joiner node becomes an online node. We refer to these steps as the node’s wake-up procedure.

\begin{definition}[Node Awakening]
\label{def:awakening}
Let honest node $i$ be a late joiner that awakens at time $t_w$. 
Node $i$ waits until time $t_{on}>t_w$ such that node $i$ does not receive any new blocks between $t_{on}-6\Delta$ and $t_{on}$. If there are two conflicting blocks $\B,\B'\in\view_i^{t_{on}}$  that are both $k$-deep in some longest chain in $\view_i^{t_{on}}$, then node $i$ halts. 
Otherwise, node $i$ certifies and finalizes all blocks that are  $k$-deep in a longest chain in view, and becomes an online honest node. 
\end{definition}

The following result establishes the protocol’s consistency guarantee for late joiners. The Stubborn Nakamoto protocol requires an additional assumption to ensure consistency for late joiners: namely, a late joiner must rely on at least one online honest node to maintain consistency with blocks that were finalized before the late joiner woke up.

\begin{theorem}
\label{thm:consis_late}
Suppose that node $i$ is a late joiner that wakes up according to \cref{def:awakening}, and there is some honest node $j$ that is online throughout and did not halt by time $t_{on}$. Then consistency holds for node $i$ even in the presence of an adversary with unbounded computational power and any number of corrupted nodes. 
\end{theorem}

\begin{remark}
\cref{thm:consis_late} does not hold under the protocol modification outlined in \cref{rem:ignore_conf_blocks}. The reason is that while online nodes can detect and ignore the attack chain described in \cref{sec:bitcoin-rental}, late joiners lack the ability to distinguish between the attack chain and the honest chain. In other words, late joiners do not have sufficient information to select the correct chain after online nodes ignore an attack chain.
These challenges are addressed by the recovery oracle, introduced in \cref{sec:SN_recovery}.   
\end{remark}

\begin{proof}
Suppose that $i$ is a late joiner, and became online at time $t_{on}$. Let $j$ be an honest node that is online and did not halt by time $t_{on}$. 

First, let us show that $\fin^\ton_i=\fin^\ton_j$. 
Let $\B_j\in\fin^\ton_j$  be a block finalized by $j$. 
Let $t_j$ denote the time at which block $\B_j$ was certified by $j$  and let $\chain(\bar{\B}_j)$ denote the corresponding chain that certified $\B_j$. Because $\B_j$ is finalized by $j$ by time $\ton$, we have that $t_j\leq \ton -2\Delta$. Therefore, $\chain(\bar{\B}_j)\subseteq \view_i^\ton$. If a longest chain in $\view_i^\ton$ contains $\B_j$, it must be as at least as long as $\chain(\bar{\B}_j)\subseteq \view_i^\ton$, and therefore include $\B_j$ at least $k$-deep. This implies that $\B_j\in \fin^\ton_i$, as required. 

Otherwise, assume for the sake of contradiction that $\chain(\bar{\B}')$ is a longest chain in $\view_i^\ton$ that does not include $\B_j$. Since $\chain(\bar{\B}')$ is at least as long as $\chain(\bar{\B}_j)$, it must include some block $\B'\conflict\B$ that is at least $k$-deep in $\chain(\bar{\B}')$. We have that 
\[
\view_i^{\ton-6\Delta}\subseteq \view_j^{\ton-5\Delta}\subseteq \view_j^{\ton-\Delta}\subseteq\view_i^{\ton}\;,
\]
and therefore $\chain(\bar{\B}')$ is a longest chain in $j$'s view at time $\ton-5\Delta$. But this implies that node $j$ must halt before time $\ton$, having seen a conflicting certified block.

To show $\fin^\ton_i\subseteq\fin^\ton_j$, let $\fin^\ton_i=\chain(\B_i)$ and let  $\chain(\bar{\B}_i)$ denote a chain that certified $\B_i$.  We have that $\h{\bar{\B}_i}\geq \h{\tilde{B}}$ for all $\tilde{\B}\in\view_i^{\ton}$. Moreover, because $i$ does not halt by time $\ton$, $\B_i$ is $k$ deep in any longest chain in $\view_i^{\ton}$. 
We have that 
\[\chain(\bar{\B}_i)\subseteq\view_i^{\ton}=\view_i^{\ton-6\Delta}\subseteq \view_j^{\ton-5\Delta}\;.\]
Therefore, node $j$ certified $\B_i$ by time $\ton-5\Delta$. Because $j$ did not halt by time $\ton$, we have that $j$ finalized $\B_i$ and every block in $\chain(\B_i)$ by time $\ton$. Thus, we have that $\fin^\ton_i=\fin^\ton_j$.

\smallskip

Now, suppose that node $i$ certifies block $\B$ at time $t_1>\ton$ and finalizes it at time $t_2=t_1+2\Delta$. 
It is immediate that $\B$ does not conflict with any block certified or finalized by $i$. Suppose, for the sake of contradiction, that honest node $j$ finalizes a conflicting block $\B'$ at time $t_2'$. If $t_2'\leq \ton$, we have  $\B'\in\fin^\ton_j=\fin^\ton_i$ by the first part of the proof, and we have a contradiction.

If $t_2' > \ton$, it must be that $j$ certifies the block $\B'$ at time $t_1'>\ton-\Delta$. Let $\chain(\bar{\B}')$ be the corresponding chain in $j$'s view at time $t_1'$. We have that $\chain(\bar{\B}')$ must be longer than any chain in $j$'s view at time $t_1'$, and therefore it is longer than any chain in $\view_i^{\ton}$. Node $i$ receives $\chain(\bar{\B}')$ by time $\tau\leq t_1'+\Delta$. Because $\view_i^{\tau-2\Delta}\subseteq \view_j^{\tau-\Delta} \subseteq \view_j^{t_1'}$, we have that $\B'\in\PC_i^\tau(2\Delta)$. Since node $i$ finalized block $\B$, we must have $\tau>t_2$. 

Let $\chain(\bar{\B})$ be the chain in $i$ view that certified $\B$ at time $t_1$. Again, we have that $\chain(\bar{\B})$ is longer than any chain in $\view_i^{\ton}$. 
Node $j$ receives $\chain(\bar{\B})$ by time $\nu\leq t_1+\Delta$. Because $\view_j^{\nu-2\Delta}\subseteq \view_i^{\nu-\Delta} \subseteq \view_i^{t_1}$, we have that $\B\in\PC_j^\nu(2\Delta)$. But this implies that $ t_1+\Delta \geq \nu > t_1'+2\Delta$ and $ t_1'+\Delta \geq \tau > t_1+2\Delta$, and we have reached a contradiction. 


\end{proof}

When the honest majority assumption holds, late joiners can reliably identify previously finalized blocks that must be in the longest chain in their view. The following theorem provides the protocol's guarantees for late joiners in the absence of an attack.

\begin{theorem}    
If the honest majority assumption holds, Stubborn Nakamoto satisfies consistency and liveness for online nodes and late joiners. 
\end{theorem}

\begin{proof}
The proof follows from \cref{thm:liveness-stubborn} together with the observation that if consistency and liveness hold for online nodes, then Stubborn Nakamoto and the standard Nakamoto protocol generate identical executions for late joiners (in the sense that any honest node maintains the same finalized ledger). 
\end{proof}

\jacob{trivial to replace one online node throughout with a sequence of nodes that come and go. But we need the sequence.}

\jacob{need to know if halted. There's a stronger impossibility, we'll talk about it in recovery}

\subsection{Recovery}
\label{sec:SN_recovery}
\jacob{Of course we can do everything if the community is strong enough. But we want it to be weak, easily inspectable, and less powerful. Used only off-path. }

We turn to consider the recovery of Stubborn Nakamoto after an attack. An attacker with sufficient resources can cause nodes to halt and the protocol to lose liveness. But suppose that the attacker ends their efforts after some time, and the protocol regains honest majority. Can the protocol recover and regain liveness while maintaining consistency? In particular, can the protocol ensure that any block finalized by a node before or during the attack does not conflict with new blocks that are finalized after recovery?

\jacob{
\textbf{TALK ABOUT THE COMMUNITY and SOCIAL CONSENSUS late joiners need access to a bunch of stuff, like an IP address...} }

\jacob{
Good properties even when halting. 
Can the protocol force the attack to pay an ongoing cost to prevent liveness?
}

\jacob{discuss guarantees for late joiners}

\jacob{
\textbf{discuss the remark about ignoring stubborn blocks}

Some caveats/details: if online nodes ignore an attack, late joiners need to use the oracle to learn which form they should join. Practically, it's asking if late joiners can identify whether they are joining Bitcoin or Bitcoin Cash. 

To ignore conflicts, need miners to aim to extend the longest chain consistent with $\fin$
}

We show that the Stubborn Nakamoto protocol can recover with external assistance that we model as a \emph{recovery oracle} and that such external assistance is necessary. The recovery oracle models a community response to the attack, allowing us to evaluate the necessary capabilities required from the community. Any protocol requires some community action to establish agreement on its rules. In particular, the Nakamoto protocol relies on an external agreement on a genesis block.\footnote{That is, the participants of the protocol must achieve agreement on the genesis block by means that are out of protocol.} We show that Stubborn Nakamoto can recover from an attack with a new genesis block provided by the recovery oracle. 

The following defines a recovery oracle that enables the recovery of the Stubborn Nakamoto protocol. This definition formalizes the exact function the community (or another external source) must provide to the protocol. The recovery oracle is practical only if its capabilities are limited to those the community can plausibly implement. While we do not formally model the community's capabilities to achieve consensus, we bear in mind that the community must have been able to initiate the protocol and provide the genesis block (potentially through a costly process) but is unlikely to maintain real-time knowledge of every honest node’s local view. Therefore, our definition of the recovery oracle relies solely on having access to the view of a single honest node. 

\jacob{want to tell if everyone halted}

\begin{definition}[Recovery Oracle]
A call to a \emph{recovery oracle} at time $t$ updates the protocol's setup by creating a new genesis block $\B_G$, which extends all clear-certified blocks in the view of some arbitrarily selected honest node. The call also awakes honest nodes that halted. 
\end{definition}

In other words, a call to the recovery oracle restarts the protocol in a similar manner to the initial launch of the Nakamoto protocol: the call announces a genesis block that any future chain must extend. All honest nodes ignore blocks that do not extend the new genesis block. The recovery oracle requires some contemporary knowledge of the state of honest nodes running the protocol, as it needs to maintain consistency with all finalized blocks. However, the recovery oracle is only required to know the state of a single (arbitrarily selected) honest node. Thus, we argue that this recovery oracle can be plausibly implemented by the community.\footnote{The Stubborn Nakamoto protocol is specifically designed to allow a recovery oracle that uses only limited capabilities (that the community can plausibly implement); see \cref{lem:recovery_consistency} and the discussion below.}

\begin{theorem}[Consistency and liveness after recovery]
\label{thm:recovery}
Under synchrony, consider Stubborn Nakomoto with a call to the recovery oracle at time $T$. Suppose that the network-aware honest majority assumption holds from time $T$ onwards. Then, Stubborn Nakamoto satisfies consistency at any time and liveness after time $T$. 
\end{theorem}

An attacker with sufficient resources can cause the Stubborn Nakamoto protocol to lose liveness. Even after a call to the recovery oracle, the attacker can repeat its attack to cause the protocol to lose liveness again. Given the impossibility results mentioned above, such a loss of liveness is unavoidable for any protocol that maintains consistency. 
However, it is costly for the attacker to continuously violate liveness: each liveness violation requires the attacker to create an attack chain that certifies or potentially certifies a conflicting block, and all blocks in the attack chain are discarded after the call to the recovery oracle. \cref{thm:recovery} guarantees that when the attacker stops and Stubborn Nakamoto regains an honest majority, Stubborn Nakamoto recovers liveness (while maintaining consistency throughout). 

\begin{remark}
\cref{thm:recovery} holds under the protocol modification described in \cref{rem:ignore_conf_blocks}. If online nodes detect an attack chain and ignore it without halting—while invoking the recovery oracle—then late joiners will be directed to the correct ledger by the recovery oracle, ensuring consistency. 
\end{remark}

The following lemma provides the key argument for the consistency of Stubborn Nakamoto after recovery. A genesis block establishes agreement on an initial ledger. \cref{lem:recovery_consistency} implies that the recovery oracle establishes a ledger that maintains consistency with any transactions finalized by honest nodes before recovery.

\begin{lem}[Consistency guarantees for recovery]
\label{lem:recovery_consistency}
If some honest node $i$ finalizes block $\B$, then every honest node $j$ clear-certifies block $\B$. If some honest node $i$ finalizes block $\B$ and does not halt, then all honest nodes finalize block $\B$.  
\end{lem}

\begin{proof}
Suppose that honest node $i$ certified block $\B$ at time $t_1$ and finalized block $\B$ at time $t_2=t_1+2\Delta$. Let $\chain(\bar{\B})$ be a logest chain in the view of node $i$ at time $t$ such that $\B$ is at least $k$-deep in $\chain(\bar{\B})$. 

Suppose for the sake of contradiction that some honest node $j$ does not clear-certify $\B$. Node $j$ receives $\chain(\bar{\B})$ by time $t+\Delta$. If $\chain(\bar{\B})$ (or an extension of it) is a longest chain in the view of node $j$, then it must be that there is some block $\B'$ such that $\B'\conflict\B$ and $\B'\in\C_j^{t+\Delta}$. But then we have by \cref{lem:certify_echo} that $\B'\in\PC_i^{t+2\Delta}(2\Delta)$, which is a contradiction to $i$ finalizing $\B$. 

Suppose then that there is a longer chain $\chain(\bar{\B}')$ in the view of node $j$ at time $t+\Delta$. Suppose that block $\B$ does not appear $k$-deep in $\chain(\bar{\B}')$ (otherwise, we reach a contradiction using the previous argument). Then there must be a block $\B'$ that is at least $k$-deep in $\chain(\bar{\B}')$ and $\B'\conflict\B$. Again, we have $\B'\in\C_j^{t+\Delta}$, leading to a contradiction. 

\smallskip
For the second part of the proof, suppose that honest node $i$ certified block $\B$ at time $t_1$, finalized block $\B$ at time $t_2=t_1+2\Delta$, and did not halt at time $t_3=t_1+4\Delta$. By the previous part of the proof, every honest node $j$ clear-certified block $\B$. Suppose for the sake of contradiction that some node $j$ did not finalize block $\B$. Observe that node $j$ certified block $\B$ by time $t_1+\Delta$, and therefore it must be that node $j$ halts without finalizing $\B$ at some time $t'\leq t_1+3\Delta$. But this implies that there exists some $\B'\in\PC_j^{t'}(2\Delta)$ such that $\B'\conflict\B$, and by \cref{lem:certify_echo} we have that $\B'\in\PC_i^{t'+\Delta}(4\Delta)\subseteq\PC_i^{t_3}(4\Delta)$ in contradiction to $i$ not halting at time $t_3$.




\end{proof}

The proof of \cref{thm:recovery} follows from the lemma. A call to the recovery oracle maintains consistency and helps the protocol regain liveness when the attacker no longer controls a majority of the nodes. We remark that the theorem holds even when there are arbitrarily many calls to the recovery oracle. 

\begin{proof}[Proof of \cref{thm:recovery}]
By \cref{thm:liveness-stubborn}, the protocol satisfies liveness after it recovers. By \cref{thm:consistency-stubborn}, if nodes $i,j$ finalize blocks $\B,\B'$ of the same height before recovery, then $\B=\B'$. Likewise, by \cref{thm:consistency-stubborn}, if nodes $i,j$ finalize blocks $\B,\B'$ of the same height after recovery, then $\B=\B'$. Suppose that node $i$ finalizes block $\B$ before recovery. By \cref{lem:recovery_consistency}, every honest node clear-certified block $\B$. Therefore, after recovery, the block $\B$ is part of the chain of the new genesis block $\B_G$ and is finalized by any honest node $j$. 
\end{proof}

The Stubborn Nakamoto with a recovery oracle provides strong guarantees to users. Consistency is always guaranteed, and protocol participants cannot be deceived. An attacker can disrupt liveness, but only temporarily. An attacker can force the community to take potentially costly actions to implement the recovery oracle, but attacks are costly for the attacker (who needs to create blocks that are discarded). Because the existence of a recovery oracle limits the harm an attacker can cause, the recovery oracle may completely deter attackers without ever being used.

The protocol relies on the recovery oracle, but the recovery oracle's influence over the system is highly constrained. First, this dependence is necessary only after the protocol is attacked. Second, any honest node can assume the role of the recovery oracle. Third, the ambiguity that the recovery oracle addresses is minimal, as the protocol ensures agreement on all blocks except those finalized within a brief time window.

\subsection{Necessity of the recovery oracle}
\label{sec:oracle_necesity}

\jacob{
-- Put the necessity of synchrony here?
}

\jacob{
We don't want to rely on "community" for everything, but maybe it's fine if they only need to do small "off path" stuff that is easy to verify? 
}

Can an alternative protocol design achieve the same properties without reliance on an external recovery oracle? Can the functionality of the recovery oracle be provided within the protocol? This section shows that the answers to these questions are negative.

The recovery oracle serves several functions. It resolves some potential ambiguity among online nodes.\footnote{Under corrupt majority, it is possible that some nodes finalize block $\B$ and other nodes halt without finalizing $\B$, and it is possible that some honest nodes certified block $\B$ and other honest nodes certified block $\B'\conflict \B$ (and no node finalizes either block $\B$ or block $\B'$).} It resolves potential ambiguity for late-joining nodes. It discards previous conflicting blocks by providing a new genesis block. Last, it informs nodes that halted that they should resume being active. 

It may be possible to implement many of these roles within the protocol. Still, without an external recovery oracle, no protocol can guarantee that late-joining nodes adopt the correct ledger, potentially violating consistency for late joiners. 

As an illustration, consider a late-joining node $i$ that attempts to join Bitcoin and become a miner. Several similar blockchains use Nakamoto consensus (for example, Dogecoin and Litecoin), and node $i$ must determine which of the chains is the Bitcoin blockchain. Formally, nodes are able to identify the Bitcoin blockchain by using the genesis block provided to it from the protocol's setup. 

However, there are also different blockchains that share Bitcoin's genesis block and its blockchain up to a certain block. For example, Bitcoin Cash was created by a so-called ``hard fork" that split off from Bitcoin and created a new blockchain. The initial portions of the ledgers of Bitcoin and Bitcoin Cash are identical, but they are independent systems, and the later parts of their ledgers diverge. Node $i$ needs more recent information to determine whether it connects to Bitcoin or Bitcoin Cash, as these are separate blockchains that share the same initial genesis block.\footnote{The hard fork that created Bitcoin Cash included technical changes to the protocol. Bitcoin itself also adopted some technical changes at a later time. We assume that node $i$ arrives after both protocols have made these technical changes, and node $i$ must rely on an external resource to identify the correct chain.} 
In particular, the total hash rate of each chain cannot provide the required identification, as node $i$ may be interested in connecting to the chain with a lower hash rate. 
The recovery oracle allows us to capture the fact that, in practice, users and nodes are able to access resources (external to the protocol) that can identify the two chains. 

More generally, the following impossibility result asserts the necessity of the recovery oracle. 

\begin{theorem}[Necessity of the Recovery Oracle]
\label{thm:oracle_necessity}
Even under synchrony, there is no permissionless protocol\todo{needs to be defined in Section 2 or 3} that allows late joiners, always satisfies consistency, and for some $\rho>0$ and $\tau\geq0$ satisfies liveness after time $T+\tau$ if from time $T$ onwards at least $1-\rho$ of the nodes are honest. 
\end{theorem}

The intuition for the proof is that under any permissionless protocol, the attacker can create a parallel copy of the ledger. We show that an attacker can simulate the execution to the point of recovery and bring a late-joining node to adopt a parallel copy that creates a consistency violation. Conceptually, such an attack is similar to a phishing attack which misdirects users to a malicious copy of a website. The attacks on Bitcoin discussed in \cref{sec:Bitcoin_incentives} can be seen as instances of this general attack.

\begin{proof}
For the sake of contradiction, suppose that such a protocol exists for some $\tau\geq0$ and $\rho>0$. We show that an attacker can violate consistency for some late joiners.

By liveness, there is an execution in which all nodes are honest and some honest node $i$ finalizes a ledger $\ledger$ at some time $t$. We can extend this execution by having the attacker create a conflicting ledger $\ledger'$ and send it to all nodes at time $t+1$. To do so, the attacker runs a parallel copy of the protocol by running $\lceil 1/\rho \rceil$ corrupt nodes for each honest node, where each corrupt node runs the protocol except for not communicating with honest nodes and ignoring messages from honest nodes. This allows the attacker to create a conflicting ledger $\ledger'$ that is finalized by the corrupt nodes. 

Because the protocol is permissionless and the corrupt nodes simulate a different possible execution of the protocol, the ledgers $\ledger,\ledger'$ are indistinguishable for late joiners. 
That is, a late joiner cannot know which ledger, $\ledger$ or $\ledger'$, has been created by corrupt nodes that do not communicate and ignore messages.

The attacker continues to run $\lceil 1/\rho \rceil$ corrupt nodes for each honest node, where from time $T=t+1$ onwards, each corrupt node runs the protocol wholly and fully communicates with honest nodes (except that all the corrupt nodes have finalized $\ledger'$ instead of $\ledger$). At time $T$, a late joiner joins the protocol. Observe that for the late joiner, this execution is indistinguishable from an execution in which the corrupt and honest nodes swap roles (that is, honest nodes finalize $\ledger'$ and corrupt nodes finalize $\ledger$) and at most a fraction $1/(1+1/\rho) < \rho$ are corrupt. Thus, at time $T+\tau$, a late joiner believes the protocol regains liveness and finalizes a ledger that extends $\ledger'$, violating consistency.  
\end{proof}

\jacob{
Comments: 
A general version of the double spend attack under Nakamoto. Nodes that are online will not be deceived, but late joiners cannot distinguish. Therefore, we must do a halt and ask for a recovery oracle. 

It is important that nodes trigger a recovery. Even if there is a fork that is ignored by online nodes of SN, we need a call to the recovery oracle to ensure that late joiners know which copy to join. 

If a call to the Recovery Oracle is called when the protocol losses honest majority, then it is safe for the protocol to ignore short fork that are not the longest chain. Although a late joiner cannot tell whether a fork was the longest chain at the time it was mined, it will be able to see the absence of a call to the recovery oracle. 

The proof is basically an extension of the "split mind" proof for the impossibility without honest majority. 
}




\jacob{Cost of doing a liveness attack. The threat of recovery can mean we don't actually need to recover.}

%% file: 5_PoS.tex
\section{Partial Synchrony and Proof-of-Stake} 
\label{sec:PoS}


In the years since \cite{satoshi-bitcoin} showed the possibility of permissionless consensus, there has been extensive work exploring protocol designs. 
An influential alternative class of protocols is commonly known as Proof-of-Stake (PoS). This class includes a wide range of protocols, which differ in their approach to consensus and the extent to which nodes can freely enter and exit.\footnote{See \cite{lewis2023permissionless} for a classification of the level of permissionlessness of different PoS protocols.} The distinctive property of PoS protocols is the use of identities and stakes: each protocol node has an identifier and an associated account with a balance that can be controlled by the protocol (commonly called the \emph{stake}).

Establishing node identifiers is helpful for both the consensus protocol and the incentives within the protocol. Identifiers allow the use of classical consensus approaches (for example, the Hotstuff protocol \citep{yin2019hotstuff}), which require nodes to sign messages cryptographically. Signed messages can provide evidence of a node's actions, potentially allowing the protocol to identify faulty nodes (see \citealt{bftforensics,accountabilitydilemma}) and punish them by reducing the balances in the node's associated accounts (commonly called \emph{slashing}). 

Our framework offers two observations on PoS protocols. 
First, PoS protocols can deliver better economic security under weaker communication requirements. 
Second, we present a variation of \cref{thm:oracle_necessity} for PoS protocols.




\subsection{Economic security for a partitioned merchant}

The ability to penalize nodes is particularly beneficial when communication is not synchronous.\footnote{For example, a payment system may want to allow a merchant to accept payment even if they are offline (e.g., a farmers market). However, even traditional payment services may find it challenging to secure offline merchants. For example, the payment processor Square states on its website:   
\begin{quote}
    ``There is additional risk when accepting payments offline. Square is not responsible for any loss due to declined cards or expired payments taken while offline.''
\end{quote} 
(\url{https://squareup.com/help/us/en/article/7777-process-card-payments-with-offline-mode}, accessed July 2024)
}
Consider an attacker attempting to deceive a merchant whose node is partitioned from the network. The partitioned merchant's node follows the protocol but only receives messages from the attacker. 
The merchant's node does not receive messages from other honest nodes in the network. Moreover, assume the merchant is unaware that it is partitioned (i.e., it cannot detect the lack of communication and halt). Assume that the merchant remains partitioned for the duration of the attack but eventually reconnects to the network.\footnote{Such partitioned merchant is possible under both the asynchronous communication model and the partial synchrony communication model \citep{dwork1988consensus}, as the attacker may delay the delivery of messages (both to and from the merchant) for an arbitrary (but finite) length of time.} 

Define the economic security for a partitioned merchant as the attacker's minimal cost of inflicting a consistency violation on a partitioned merchant. 

Observe that, under any permissionless protocol, the attacker can deceive a partitioned merchant by creating a conflicting copy of the ledger. 
To do so, the attacker simulates a parallel execution of the protocol that is indistinguishable from the protocol's correct operation to the partitioned merchant (recall that the merchant cannot communicate with honest nodes and hence cannot detect the existence of another, conflicting ledger). 
This observation implies an immediate upper bound on the economic security for a partitioned merchant.

\begin{remark}
Under a permissionless protocol, the economic security for a partitioned merchant is upper bounded by the cost of simulating an execution of the protocol that leads the merchant's node to finalize a transaction, plus the potential costs of punishments incurred when the merchant reconnects and communicates with other honest nodes.
\end{remark}

Under both Nakamoto and Stubborn Nakamoto, the attacker needs to create a conflicting chain of $k$ blocks to get the partitioned merchant to finalize a transaction and violate consistency. 
In the rental model, the attacker's cost of creating $k$ blocks is $k\cdot p_\B$. The two protocols differ in the response of honest nodes after the partition ends. Under Nakamoto, the attacker can induce all honest nodes to adopt its conflicting ledger, enabling the attacker to collect block rewards and recoup its costs (as described in \cref{sec:bitcoin-rental}). By contrast, under Stubborn Nakamoto, the net cost to the attacker remains $k\cdot p_\B$. The attacker can deceive a partitioned merchant, but conflicting blocks cannot become part of the consensus chain. Note that under Stubborn Nakamoto, nodes are anonymous, and there is no additional punishment for creating a conflicting chain. 

Under many PoS protocols, the cost of simulating a (conflicting) execution of the protocol is negligible. Instead, the protocol may be able to detect and penalize nodes that create a conflicting ledger (for example, by witnessing the node's signature on the conflicting ledger). In particular, if the protocol can detect nodes that have deviated from the protocol, it may punish only the deviating nodes and thus avoid the tragedy of the commons\footnote{To completely avoid the tragedy of the commons, the protocol should punish deviating nodes even when there is no successful attack. } described in \cref{sec:bitcoin_bribe}. 

To establish an upper bound, consider a hypothetical PoS protocol that can perfectly identify all the nodes that create a conflicting ledger (and assume that the cost of protocol execution is 0). The maximal possible punishment that can be imposed on a node is the slashing of its entire stake (clearing the balance in the node's associated account). Thus, an upper bound for the attacker's total cost is the entire stake of all the nodes that created the conflicting ledger. In other words, the economic security for a partitioned merchant is bounded by the total amount of stake required to create an execution of the protocol that finalizes a transaction.

A simple comparison suggests that punishment in PoS can lead to much greater economic security for a partitioned merchant than the cost of execution in Stubborn Nakamoto. If the Stubborn Nakamoto protocol spends $p_\B$ per block, under the rental model, it achieves economic security for a partitioned merchant of $k\cdot p_\B$. A PoS protocol can potentially achieve greater economic security by requiring nodes to stake. 
Suppose that the cost of executing the protocol is negligible, and the protocol only needs to compensate nodes for forgone interest. In that case, block rewards should equal forgone interest on the total stake, and nodes provide a total stake of $p_\B / r$, where $r$ is the interest rate per block.\footnote{The protocol can avoid these costs if it can invest the staked funds. We thank Niels Gormsen for this observation.} If the attacker suffers losses equal to a fraction $\lambda$ of the total stake, the attacker's cost is $\lambda p_\B / r$. 

However, note that PoS protocols face several challenges that are omitted from the above discussion.  
It may be challenging to detect deviating nodes, and the protocol may detect only some of the deviating nodes \citep{bftforensics,accountabilitydilemma}. If the protocol mistakenly slashes honest nodes, it must compensate all nodes for this potential harm. Moreover, the attacker may cause further deviations from the protocol to prevent the protocol from enacting the punishment \citep{budish2024economic}.

\subsection{Necessity of Recovery Oracles}
We provide a version of the impossibility result given in \cref{thm:oracle_necessity} for PoS protocols. Consider a PoS protocol that guarantees consistency. Concurrent work of \cite{sridhar2023better} shows the feasibility of such PoS protocols under synchronous communication. But no protocol can always guarantee consistency and liveness under synchronous communication with late joiners \citep{pass2017rethinking,lewis2023permissionless}.\footnote{If all nodes are online and communication is synchronous, then the Dolev--Strong protocol guarantees consistency and liveness even without honest majority. But honest majority is necessary for the consistency of the Dolev--Strong protocol when late joiners are allowed.} Can a PoS protocol only lose liveness temporarily, halting progress while there is no honest majority, and regain liveness once honest majority is restored? 

By \cref{thm:oracle_necessity}, a recovery oracle is necessary if the protocol is permissionless. 
In practice, many PoS protocols require that active nodes approve late-joining nodes.\footnote{For example, a node may be required to issue a transaction that establishes its stake before participating in the protocol. If such transactions are written to the ledger, and active nodes must process them (and can censor them), late-joining nodes will require approval from active nodes.} 
This required approval can help secure the protocol by limiting the attacker's ability to simulate a conflicting protocol execution. However, such a limitation is helpful only if it introduces a different problem: the protocol can get stuck if no late-joining nodes are allowed to enter. We state this tension in the following corollary. 

\begin{corollary}
\label{cor:PoS_necessity}
Consider a protocol that, under synchrony, allows late joiners, always satisfies consistency, and for some $\rho>0$ and $\tau\geq0$ satisfies liveness after time $T+\tau$ if from time $T$ onwards at least a fraction $1-\rho$ of the nodes are honest. 

Then there exists some  $\gamma\leq 1$ such that if a fraction $\gamma$ of the nodes crash, the protocol loses liveness forever. 
\end{corollary}

The proof follows a similar argument to the proof of \cref{thm:oracle_necessity}. 

\begin{proof}
Suppose, for the sake of contradiction, that such a protocol exists and regains liveness at some time after all nodes crash at time $t$. Then, it must be possible for late-joining nodes to join the protocol at some time $\tau>t$ despite all previous nodes crashing. Similarly to the proof of \cref{thm:oracle_necessity}, the attacker can exploit this possibility to create a conflicting ledger $\ledger'$ by launching a separate collection of nodes that follow the protocol, except that they run as if all nodes crashed at time $t$. 

Alternatively, suppose that the attacker controls all existing nodes at time $t$. The attacker continues running the nodes but makes it appear as if all nodes crashed. New late-joining honest nodes enter the protocol and generate the ledger $\ledger$. Because a late joiner cannot distinguish between the two scenarios, the protocol cannot guarantee consistency for late joiners.
\end{proof}

\cref{cor:PoS_necessity} and \cref{thm:oracle_necessity} show that any protocol must require some external help under some scenarios: either to recover from attacks or to recover after all active nodes halt. In other words, some dependency on the community is unavoidable. Still, protocols can minimize this dependency by relying on external help only in rare scenarios and reducing the requirements needed for implementing such external help.

\jacob{Points to make:
\begin{itemize}
    \item Either we get the same "split brain" problem after recovery, or there is some restriction on entry and need external oracle if all current nodes decide to halt.
    \item Bound for partitioned client: cost of creating the log + cost of punishment
    \item Slashing should only depend on what the slashed node does (e.g., not on whether that resulted in damage to the chain), because we want to avoid tragedy of the commons. 
    \item Long range attacks can be prevented
    \item 
\end{itemize}
}

\ignore{

}

%% file: 6_Discussion.tex
\section{Discussion}
\label{sec:Discussion}

Our results demonstrate the feasibility of an open-source record-keeping service that operates without a trusted record keeper. While users may not be concerned with the internal structure of the system, they can observe its beneficial properties: it protects its users from deception, allows for free entry and exit of operators, and protects users from monopoly's harm.

Central to our design is prioritizing consistency (security) over liveness (timely ledger updates). Roughly speaking, the Stubborn Nakamoto protocol either completely avoids the consistency violations of the Nakamoto protocol or converts them into liveness violations.
We argue that financial applications should prioritize consistency over liveness. Recording a transaction in a financial ledger is beneficial only if other parties recognize that record. 
If the ledger is updated with records that are not recognized or are later invalidated, users do not receive any benefit and are exposed to deception and financial loss.
Another justification for prioritizing consistency is that an attacker can clearly benefit from consistency violations that enable deception. By contrast, liveness violations do not offer the attacker such direct gains. 


Many questions about the design of payment systems are beyond the scope of this paper. The Stubborn Nakamoto protocol provides a general ledger, enabling users to record arbitrary data. Bitcoin is one example of such a payment system. The technology allows for many others. For example, the ledger could be used to record balances in USD or document land ownership. The system may allow users to be  
pseudoanonymous, or may requier that every user is fully identifiable. Such design choices raise a significant host of issues, including privacy concerns and anti-money laundering challenges. For a discussion of such challenges, see, for example, \cite{rogoff2017curse} and \cite{makarov2022cryptocurrencies}.

Finally, one significant benefit of open-source technology is its inherent openness and modularity. 
Much like how anyone can join a permissionless protocol as a node, open-source frameworks enable developers to build upon existing services. 
This approach encourages innovation and competition in the market for higher-level services. 
For instance, while foundational Internet protocols such as TCP/IP may not be of much use to end users alone, they establish a crucial infrastructure enabling user-friendly commercial applications. 
Similarly, even permissionless protocols that provide limited functionality can be beneficial as foundational rails that allow firms to build more complex services, fostering a competitive market where diverse and innovative services can thrive.


%% file: A1_Appendix.tex
\section{Distributed Consensus Formal Execution Model}
\label{sec:appendix-model}

The proof of \cref{thm:liveness-stubborn} uses a similar approach to the proof of consistency and liveness of the Nakamoto protocol of \cite{pass2017analysis,shitextbook}. The proof requires a few additional definitions and utilizes several stochastic bounds that were proved in earlier works \citep{pass2017analysis,shitextbook}. We state these here.

\subsection{Execution model}
We introduce some notation regarding the protocol's execution model in order to formally state the conditions under which Nakamoto has provable guarantees.

We assume that $n$ nodes participate in the consensus protocol, all with equal mining power.\footnote{A node with larger mining power can be viewed as multiple nodes.}
Time is discrete and divided into rounds. Each round corresponds to one calculation of a hash by a miner. Let $p$ be the probability that a single node finds a valid block in a round, where the probability $p$ is related to the mining difficulty parameter. We use the notation $\rho \in [0, 1)$ to denote the fraction of nodes controlled by the adversary.

\subsection{Definitions}
Because the Nakamoto consensus protocol requires randomness, a small probability of failure may be inevitable. The formal definition of liveness for the Nakamoto and Stubborn Nakamoto protocols bounds the failure probability by a security parameter $\lambda$. The protocol can tune its security parameter $\lambda$ by selecting a required block depth $k$, such that $k=\omega(\log \lambda)$. 

For the sake of simplicity, we assume that an honest node that mines a block can include all the outstanding valid transactions (not included in the longest chain) given its local view. Let $T_{\rm conf} = T_{\rm conf}(\lambda, \Delta, n, \rho)$ be a function of the security parameter $\lambda$, network delay $\Delta$, number of nodes $n$, and fraction of adversary nodes $\rho$. We call $T_{\rm conf}$ the \emph{confirmation time}.\todo{poly-length execution}

\begin{definition}[$T_{\rm conf}$-Liveness]
The protocol satisfies $T_{\rm conf}$-Liveness if there is some negligible\footnote{
We say that ${\sf negl}:\mathbb{R}\to\mathbb{R}_+$ is a negligible function iff for any
fixed polynomial function $p(\lambda)$, there exists $\lambda_0$ such that for any $\lambda>\lambda_0$,
${\sf negl}(\lambda) < 1/p(\lambda)$. In other words, a negligible function is one that vanishes faster than any inverse-polynomial function.
}%
function ${\sf negl}(\cdot)$ such that with $1-{\sf negl}(\lambda)$ probability when some honest node receives a transaction ${\sf tx}$ at some time $t$, then by time $t + T_{\rm conf}$ all honest nodes' finalized ledgers include ${\sf tx}$.
\end{definition}

We can now formally state the network-aware honest majority assumption used in \cref{thm:liveness-stubborn}. 

\begin{definition}[Network-Aware Honest Majority]
Consider a polynomial length execution of Stubborn Nakamoto with a required block depth $k$ that is super-logarithmic function in the desired security parameter $\lambda$. Let $\nu = 2 p n \Delta$. The Network-Aware Honest Majority assumption is satisfied if $\nu<1/2$ and there exists some (arbitrarily small) constant $\phi \in (0, 1)$ such that 
\[(1-\rho)(1-\nu) \geq (1+\phi)\rho\;.\] 
\end{definition}

The term $(1-\rho)(1-\nu)$ can be viewed as the fraction of honest mining power discounted by the network delay $\Delta$. Therefore, this assumption requires that the honest mining power, even after accounting for network delay, exceeds the corrupt mining power by a small constant margin.

\subsection{Stochastic bounds}
\label{sec:appendix-lem}

The following stochastic bounds proved in earlier works \citep{pass2017analysis,shitextbook} are used in the proof of \cref{thm:liveness-stubborn}. We state them here for completeness. 

Define convergence opportunity in the same way as Section 17.3 of \cite{shitextbook}. Specifically, round $t$ is a convergence opportunity if a single honest block is mined in round $t$ and no honest blocks are mined in the surrounding $\Delta$ rounds before and after. Given a run of Nakamoto or Stubborn Nakamoto, let ${\bf C}[t:t']$ denote the number of convergence opportunities between $t$ and $t'$, and let ${\bf A}[t:t']$ denote the number of adversarial blocks\footnote{That is, blocks mined by nodes that are not honest nodes.} mined between $t$ and $t'$. Let $\alpha := p\cdot(1-\rho)n$ denote the expected number of honest nodes that mine a block in a given round, and let $\beta := p\cdot \rho n$ denote the expected number of corrupt nodes that mine a block in each round.

\begin{lemma}[Lower bound on convergence opportunities~\citep{pass2017analysis,shitextbook}]
\label{lem:convopp}
For any positive constant $\varepsilon>0$ and any $k$ that is a super-logarithmic function in $\lambda$, except with negligibly small in $\lambda$ probability over the choice of the execution, 
the following holds:
for any $t_0, t_1 \geq 0$ such that $t := t_1 - t_0 > k/\alpha$,  we have that
\[
C[t_0 : t_1] > (1 - \varepsilon)(1 - 2pn\Delta)\alpha t \;.
\]
\end{lemma}

\begin{lemma}[Upper bound on adversarially mined blocks~\citep{pass2017analysis,shitextbook}] 
\label{lem:advblock}
For any constant $0 < \varepsilon < 1$, for any $k$ that is a super-logarithmic function in $\lambda$, except with negligibly small in $\lambda$ probability over the choice of the execution, the following holds: 
for any $t_0, t_1 \geq 0$ such that $t := t_1 - t_0 > k/\beta$,  we have that
\[
A[t_0 : t_1] \leq (1 +\varepsilon)\beta t \;.
\]

for any $t \geq k/\beta$. That is,
the number of adversarially mined blocks in any $t$-sized window
is upper bounded by $(1+\epsilon) \beta t$.

\end{lemma}

\begin{lemma}[Total block upper bound~\citep{pass2017analysis,shitextbook}] 
\label{lem:totalblock}
For any positive constant $\varepsilon$  and any $k$  that is a super-logarithmic function in $\lambda$, except with negligibly small in $\lambda$ probability over the choice of the execution, the following holds: 
for any $t_0$ and $t_1$ such that $np(t_1-t_0) \geq k$, the total number of blocks successfully mined during $(t_0, t_1]$
by all nodes (honest and corrupt alike) is upper bounded by $(1 + \varepsilon)np(t_1 - t_0)$.
\end{lemma}

\subsection{Proof of Theorem \ref{thm:liveness-stubborn}}

\begin{proof}[Proof of \cref{thm:liveness-stubborn}]

Given an execution of the Stubborn Nakamoto protocol, we can construct an execution of the Nakamoto\footnote{We use the same confirmation depth $k$ as our Stubborn Nakamoto for the execution of the Nakamoto protocol.} protocol. If the two executions are identical, in the sense that any honest node maintains the same finalized ledger at any point during the run, then Stubborn Nakamoto's properties follow from those of Nakamoto. We show that under the network-aware honest majority assumption, the two executions are identical except for a negligible probability, and therefore the $T_{\rm conf}$-Liveness of Stubborn Nakamoto follows from the $T_{\rm conf}$-Liveness and consistency of Nakamoto \citep{pass2017analysis,shitextbook}. 

\smallskip
The reason the executions may differ is that an honest node $i$ may halt under Stubborn Nakamoto. 
Consider the first round $t_1$ such that some honest node $i$ certifies a block $\B$ and by round $t_1+4\Delta$ node $i$ halts having seen a $4\Delta$-potentially certified block $\B'$ that conflicts with $\B$. (That is, node $i$ halted at either steps 1, 2 or 3) 



Let $\chain(\bar{\B})$ be the chain in node $i$'s local view in round $t_1$ containing $\B$ at least $k$-deep. Let $\B'\in\PC_i^{t_1+4\Delta}(4\Delta)$ be a block that conflicts with $\B$, and let $t_1\leq t \leq t_1+4\Delta$ be the fist time at which $\B'\in\PC_i^{t}(4\Delta)$. 
Let $\chain(\bar{\B}')$ be the chain in node $i$'s local record such that $h(\bar{\B}') - h(\B') +1 \geq k$ and $\h{\bar{\B}'}\geq \h{\tilde{\B}}$ for any $\tilde{\B}\in {\view}_i^{t-4\Delta}$. That is, block $\B'$ is at least $k$-deep in the chain $\chain(\bar{\B}')$,  and the chain $\chain(\bar{\B}')$  is longer than any chain in the view of node $i$ as of $t-4\Delta$.

Let $\B_{\rm common}$ the last block shared in common by $\chain(\bar{\B})$ and $\chain(\bar{\B}')$. Suppose that $\B_{\rm common}$ is mined in round $s-1$. It holds that all blocks in $\chain(\bar{\B})$ and $\chain(\bar{\B}')$ after $\B_{\rm common}$ are mined in round $s$ or later. 
Let $\tau = t - s$. Because the chains $\chain(\bar{\B})$ and $\chain(\bar{\B}')$ differ by at least $k$ blocks, by total block upper bound (see \Cref{lem:totalblock}), it must be that $\tau > \frac{k}{2pn}$.


Observe that it must be that ${\bf C}[s:t-5\Delta] \leq {\bf A}[s: t]$. Otherwise, there would have been a block $\B$ mined by an honest node during a convergence opportunity within $[s, t-4\Delta-\Delta]$, and that block $\B$ would have to appear in both $\chain(\bar{\B})$ and $\chain(\bar{\B})$ (see Section 17.3 of \citealt{shitextbook}).
Therefore, it suffices to prove that, except with negligible  $\lambda$ probability, it must be that ${\bf C}[s:t-5\Delta] > {\bf A}[s: t]$.

Let there be some $\varepsilon, \varepsilon'\in (0,1)$. By the lower bound on convergence opportunity (see \Cref{lem:convopp}), except with negligible probability, we have that
\[
{\bf C}[s: t-5\Delta] > (1-\varepsilon) (1-2pn\Delta) \alpha (\tau -5\Delta) \; ,
\]
where $\alpha := p\cdot(1-\rho)n$ denotes the expected number of honest nodes that mine a block in a given round.
By adversarial block upper bound, (see \Cref{lem:advblock}) we have that for any positive constant $\varepsilon_a$, it holds that
\[
{\bf A}[s: t] < (1+\epsilon_a) \beta \tau \;,
\]
where $\beta = p \cdot \rho n$ denotes the expected number of corrupt nodes that mine a block in each round.

Thus, for any positive constant $\phi>0$, as long as $0 < 2pn\Delta < 0.5$, there exist sufficiently small positive constants $\varepsilon, \varepsilon'$, and $\varepsilon_1$ such that the following holds for sufficiently large $k$:
\begin{align*}
{\bf C}[s:t-5\Delta] & >  (1-\varepsilon) (1-\nu) \alpha(\tau- 5\Delta) \\
& > (1-\varepsilon_1) (1-\nu) \alpha \tau \\
& > (1-\varepsilon_1) (1+\phi) \beta \tau \\
& > (1+ \varepsilon') (1 + \frac{\phi}{2}) \beta \tau\\
& > (1+ \varepsilon') \beta \tau  
\end{align*}
where the  last inequality holds for sufficiently large $k$ because $\alpha \Delta = O(1)$ and $\alpha \tau = \Theta(k)$. 
\elaine{NOTE: this assumes that corrupt mining power is not too small}

This implies that except with negligible in $\lambda$ probability we have that $ {\bf C}[s:t-5\Delta]  > {\bf A}[s: t]$, completing the proof. 

\end{proof}

\section{Other Omitted Proofs}
\label{sec:omitted-proofs}
\begin{proof}[Proof of \cref{prop:NakaRentalAttack}]
Suppose that the transaction of interest to the attacker and merchant is in block $\B_n$. The attacker waits for the merchant to confirm the transaction, which (by liveness) occurs at some time $t$. Suppose that at time $t$ the longest chain recorded by any honest miner is $(\B_0,\dots,\B_{n-1},\B_{n},\dots, \B_{n+K} )$. 

At time $t$, the attacker mines the attack chain $(\B_0,\dots,\B_{n-1},\B'_{n},\dots, \B'_{n+L} )$ for some $L>K$ and communicates it immediately to all other miners. If the attack chain is the longest chain known to any miner, it becomes the consensus chain and is adopted by all honest miners. 

The attacker pays $D\cdot c$ in expectation per mined block. If honest miners adopt the attack chain, the attacker receives $p_\B$ per mined block. Because honest miners also pay $D\cdot c$ in expectation per mined block and are profit-maximizing, it must be that $p_\B\geq  D\cdot c$, and the attacker's net cost is zero. Finally, the probability that the attack chain of height $n+L$ becomes the longest chain can be arbitrarily close to one if the attacker computes the same number of hashes over a shorter time.
\end{proof}

\section{Examples of Profitable Consistency Violations}
\label{sec:ex_profitable_attacks}
Attackers can profit from consistency violations by deceiving and taking advantage of a counterparty. We give a few examples. 

\paragraph{Double spend}
The commonly considered double-spend attack is a form of consistency violation. In it, the attacker sends the same funds to two different parties in two different transactions, but only one of the two can be valid. If consistency holds, the receiving party can detect which transaction is valid. Conversly, if the attacker can deceive a merchant who follows the protocol into accepting a transaction that conflicts with the consensus ledger, then consistency is violated. 

\paragraph{History rewrite}
Consistency requires that any node's local ledger agrees with any previously finalized data. In particular, it implies that the ledger cannot revert finalized transactions. A dishonest attacker can gain if they can undo a finalized transaction. For example, consider a transaction that transfers ownership of an asset from the attacker to a merchant. If the price of the asset increases after the sale, the attacker can benefit from reverting the transaction and retroactively canceling the transfer of ownership.

%% file: A2_what_deters_attacks.tex
\section{Other Factors Deterring Attacks}
\label{sec:appendix_what_deters_attacks}

\jacob{
From old notes:

financing cost of attack is minor.
but all relies no fixed entry cost--with ASIC the attacks is much more
costly.

bribing just not to release block: more expensive as you still need to
create your own fork.

consider multple attackers at the same time, people have different
beliefs on which fork survives.
}

This section explores additional factors that deter attacks. Despite numerous attacks on cryptocurrencies\todo{add list} using the Nakamoto protocol, major cryptocurrencies such as Bitcoin and Ethereum have remained secure. 
The rental model and the bribery model discussed in Section \ref{sec:Bitcoin_incentives} fail to explain why attacks on these protocols are not commonplace. 
This indicates that, in practice, other factors contribute to deterring attacks. 
We discuss such factors and identify elements that can be incorporated into protocol design to enhance security.

We categorize these factors into three groups. First is the user community's reaction to an attack on the protocol. The second category encompasses economic frictions that hinder the execution of an attack. The third category includes reasons for the system's security that suggest that the system is secure because it is not in fact decentralized. Altogether, the discussion suggests that Bitcoin's economic security comes not from the cost of mining but rather from these external deterrents described below.


\paragraph{Community response}
Even if the attack is not detectable within the protocol, the community can detect an attack. Much of the literature (e.g., \citealt{budish2022economic,auer2019beyond,garratt2023fixed}) considers that an attack will trigger a decline in the cryptocurrency exchange rate. This assumption inherently relies on the public's awareness that an attack has occurred.

In practice, the community can identify the attack fork and the honest fork under the attack described in \ref{sec:bitcoin-rental}. As described in Section \ref{sec:Bitcoin_incentives}, in the view of an honest node, an attack appears to be a reconciliation after a network partition. That is, an honest node will receive the blocks in the attack fork significantly later than blocks of comparable height in the honest fork.\footnote{The attacker must keep hidden the blocks in the attack chain $(\B'_{n},\dots, \B'_{n+L} )$ until the merchant finalizes the block $\B_{n}$ (otherwise, the merchant will not finalize $\B_{n}$). Thus, an honest miner receives block $\B'_{n}$ only after receiving block $\B_{n+k}$, which it receives significantly later than block $\B_{n}$. Recall that honest nodes mine conflicting blocks $\B_{n},\B'_{n}$ only if these blocks are mined simultaneously (up to network delays).} Any honest node active during the attack can recognize the attack chain as either a deliberate attack or a highly improbable event of partition reconciliation.

The community can collectively decide to undo the attack. If the community agrees that the miners were attacked and the ledger was changed illegally, the community can direct all miners to follow with the correct honest fork and ignore the attack fork. Doing so reverts any effect of the attack, leaving the ledger unchanged by the attack. The community response poses a significant risk to the attacker, who is left to pay all costs of the attack, but without the attack having any eventual effect on the ledger. 

While community-driven interventions in blockchain networks are rare, they are not without precedent. 
A notable instance is the Ethereum community's decision to amend the Ethereum ledger in response to the DAO attack.\footnote{See, for example, \url{https://blog.ethereum.org/2016/07/20/hard-fork-completed}, accessed February 2024.} 
Such measures, however, necessitate a community that possesses both the capability and the willingness to undertake manual interventions.
For example, Ethereum Classic rejects such manual interventions.\footnote{\url{https://ethereumclassic.org/why-classic/code-is-law}}
Interestingly, Ethereum did not suffer any consistency attacks, whereas Ethereum Classic suffered from multiple attacks.\footnote{\url{https://www.coinbase.com/blog/deep-chain-reorganization-detected-on-ethereum-classic-etc}, accessed February 2024.}

Note that a community response can deter attacks even if the community can only detect an attack (without agreeing on the correct ledger that should be restored). As discussed in \cref{sec:bitcoin_bribe}, preventing the attacker from collecting the block rewards after the attack lowers the payment to attackers relative to honest miners (when there is no attack). This creates a penalty for an attack that generates some detterance.

\paragraph{Frictions }
In practice,  it may be difficult to operationalize an attack. For example, even if a bribery contract is financially beneficial for miners, it may be difficult to inform and recruit miners. Moreover, miners may not be fully profit-driven, or may 
be altruistic. This poses a risk to the attacker. A bribery attack needs to commit to paying miners who mine blocks in the attack chain. If the bribery attack fails to solicit enough mining power to cooperate, the attacker faces financial risk from paying for blocks on the attack chain only for the attack to fails. 

The analysis in Section \ref{sec:Bitcoin_incentives} assumes that block rewards are fixed and there is no difficulty adjustment. \cite{gans2024zero} show that an attacker can gain by affecting block rewards and, in particular, transaction fees. The attacker is advantaged as it can observe all the transactions processed by honest miners, and can collect all fees from transactions that are not invalidated by the attack.\footnote{This may pose a limitation on an attacker that attempts to fork a very old transaction, as the fork will invalidate all the new coins given to miners in the forked blocks and all subsequent transactions these funds are exchanged in.}  

The difficulty adjustment can also be exploited by the attacker. For example, mining difficulty increases over time as mining equipment becomes more efficient, and the attacker can benefit from forking old blocks with lower difficulty using current mining equipment.  



\paragraph{Centralized factors}
An attacker may be deterred by the threat of legal prosecution. In practice, to cash out from an attack, the attacker will need to withdraw fiat money from some exchange. However, exchanges are subject to know-your-customer (KYC)
rules that require them to know the identities of their customers.\footnote{
There are many examples of hackers returning illicitly obtained funds; for example:
\url{https://www.reuters.com/technology/defi-platform-poly-network-reports-hacking-loses-estimated-600-million-2021-08-11/}, accessed August 2024.
} 

The analysis in \cref{sec:bitcoin_bribe} of the bribery model assumes small players with infinitesimally small mining power, whose actions are non-pivotal to the outcome whether the attack succeeds or not. In practice, mining power is more concentrated in the form of mining farms and mining pools. 
For a big miner, whether it cooperates in the attack may be pivotal to whether the attack is successful. For example, consider the extreme case where the attack is guaranteed to succeed if a big miner cooperates and to fail otherwise. To bribe such a big miner, it is no longer sufficient for the attacker to offer only a slightly higher than the normal block reward; it must
also compensate for the individual cost to the big miner should the attack succeed (e.g., the loss suffered from currency devaluing).
Therefore, although mining power concentration hurts the decentralized nature of the cryptocurrency, it actually helps enhance economic security in this respect. 

We point that while these factors may be effective in practice, they call into question the claims of the system's ``decentralization''.\footnote{\cite{huberman2021monopoly} find that Bitcoin offers users protection from monopoly's harm even if there are large miners, as long as free entry of miners is possible. It is technically possible for a system to have both large pivotal miners for security and free entry of miners. However, in Nakamoto a sufficiently large miner is able to block free entry of miners.}

\paragraph{Discussion}
It is commonly believed that higher mining rewards and more mining power increase the economic security of a cryptocurrency. However, our economic analysis reveals interesting subtleties challenging this claim. Both of the economic environments we analyze suggest that higher mining rewards do not actually increase the cost of the attack. 
Overall, the connection between economic security and mining rewards may be more subtle or dependent on specific circumstances.

\ignore{
As mentioned, in the rental model and the bribery model with small miners, the cost of an attack is 0 or arbitrarily close to 0. So in practice, why don't we see consistency attacks more often?
The reason is that in practice, various other factors 
help to deter attacks.
An interesting conclusion to draw from our economic analysis is that the {\it economic security of Bitcoin is comes not from the cost of mining, but rather, from these external deterrents} mentioned below.

\begin{enumerate}
\item {\it Community hard fork.}
In practice, if an attack is identified, 
the community can apply forensic techniques to 
distinguish which is the attack fork and which is the benign fork. 
At this moment, 
the community can jointly make a decision to ignore the attack fork and go with the benign fork. In this case, the attacker may not be able to cash out the block rewards on the attack fork in time to help expense
the cost of mining the attack fork (either through rental or bribery).

\item {\it Law enforcement.}
In practice, to cash out from an attack, the attacker will need to withdraw fiat money from some exchange. However, exchanges are subject to know-your-customer (KYC)
rules that require them to know the identities of their customers. In this way, the attacker can risk legal prosecution for launching the attack.

\item {\it Mining power concentration.}
Our earlier analysis of the bribery model assumes small players with infinitesimally small mining power, whose actions are non-pivotal to the outcome whether the attack succeeds or not. In practice, mining power is more concentrated in the form of mining farms and mining pools. 
For a big player, whether it cooperates in the attack may be pivotal to whether the attack is successful. For example, consider the extreme case where the attack is guaranteed to succeed if a big miner cooperates; otherwise it will fail. To bribe such a big miner, it is no longer sufficient for the attacker to only offer slightly higher than the normal block reward; it must
also compensate for the $c_A$ part of the cost to the big miner should the attack succeed (e.g., the loss suffered from currency devaluing).
Therefore, although mining power concentration hurts the decentralized nature of the cryptocurrency, it actually helps enhance the economic security in this respect. 

\item {\it Other deterrents.}
In practice, there may also be other deterrents. For example, miners may not be fully profit-driven, or may 
be altruistic. In such cases,  a bribery attack may fail due to not being able to solicit enough mining power to cooperate. In this case, the attacker cannot use the block rewards on the attack fork to help expense the cost of the attack.
\end{enumerate}

In a two-sided market with free entry, the cost of mining a block equals
the mining reward~\cite{huberman2021monopoly}.
It is commonly believed that higher mining reward (i.e., higher cost of mining)  increases the economic security of a cryptocurrency. However, our economic analysis reveals interesting subtleties regarding this claim. Specifically, our analysis suggests that in a decentralized world with small miners, {\it higher mining reward
only increases the initial capital cost needed for an attack; it does not actually reduce the cost of the attack itself.} 
}





